\documentclass[11pt,a4paper]{article}
\usepackage{epsfig,graphicx,amsfonts,amsmath}

\usepackage{cite}
\RequirePackage{mathrsfs}
\usepackage{color}
\usepackage{epsf}
\usepackage{epstopdf}

\newlength{\dinwidth}
\newlength{\dinmargin}
\def\eq#1{{Eq.~(\ref{#1})}}
\newcommand{\Le}{\left(}
\newcommand{\Ra}{\right)}

\newcommand{\beq}{\begin{equation}}
\newcommand{\eeq}{\end{equation}}
\newcommand{\beqar}{\begin{eqnarray}}
\newcommand{\eeqar}{\end{eqnarray}}
\newcommand{\D}{\partial}
\newcommand{\g}{{\rm g}}

\newcommand{\tv}{\textsl{v}}

\newcommand{\B}{{\cal B}}
\newcommand{\T}{{\cal T}}

\date{}
\begin{document}

\title {{~}\\
{\Large \bf On correlators of Reggeon fields and operators of  Wilson lines in high energy QCD  }}
\author{
{~}\\
{\large
S.~Bondarenko$^{(1) }$,
S.~Pozdnyakov$^{(1) }$
}\\[7mm]
{\it\normalsize  $^{(1) }$ Physics Department, Ariel University, Ariel 40700, Israel}\\
}

\maketitle
\thispagestyle{empty}

\begin{abstract}
  In this note we derive Dyson-Schwinger hierarchy of the equations for the correlators of reggeized gluon fields in the framework of Lipatov's
high energy QCD effective action formalism, \cite{LipatovEff,Our1,Our2,Our3,Our4}.
The  explicit perturbative expressions for the correlators till correlator of four Reggeon fields inclusively are obtained and different perturbative
schemes for the solutions of the equation for the two-field correlator are discussed.
A correspondence between the correlators of reggeized gluon fields and Wilson line
operators of longitudinal gluon field is established with the help of \cite{Our4} paper results.
The connection between the JIMWLK-Balitsky formalism and Lipatov's effective action approach
and applications of the obtained results are also discussed.
\end{abstract}

\section{Introduction}

$\,\,\,\,\,\,$ The action for interaction of reggeized gluons introduced in \cite{LipatovEff},  see also \cite{Our1,Our2,Our3,Our4},
describes quasi-elastic amplitudes of high-energy scattering processes
in the  multi-Regge kinematics. The applications of this action for the description of  high energy processes
and calculation of sub-leading, unitarizing  corrections to the
amplitudes and production vertices can be found in \cite{EffAct}.
The generalization of the formalism for a case of production amplitudes was considered in \cite{Our3} as well,
where the prescription of the calculation of $S$-matrix elements of the different processes was given following to the approach of \cite{Faddeev}.
This effective action formalism, based on the reggeized gluons as main degrees of freedom, see \cite{BFKL}, can be considered as reformulation of the RFT (Regge Field Theory) calculus introduced
in \cite{Gribov} for the case of high energy QCD.
It was underlined in \cite{LipatovEff} that the main purposes of the approach is the construction of the
$S$-matrix unitarity in the direct and crossing channels of the scattering processes through the multi-Reggeon dynamics described with the use of the vertices  of multi-Reggeon
interactions, see similar approaches in \cite{BKP, GLR, BK, TripleV}.
Therefore, one of the main goals of the RFT construction is a calculation of the scattering amplitudes in the formalism which account
unitarity corrections due a change of the number of reggeized gluons in the scattering processes.
The main ingredients of this scheme are correlation functions of the Reggeons fields
which describes these non-linear corrections to the amplitudes.
So far, the problem of hierarchy of the correlators of interests was not formulated precisely in the framework of BFKL approach
but similar pproblem
was consiered in CGC (Color Glass Condensate), BK and Balitsky-JIMWLK approaches to high energy scattering, see
\cite{Venug,BK,Kovner,Hatta1} for degrees of freedom different from the reggeized gluons.

  It was demonstrated in \cite{Our1} that an application of a shock wave and large $N_{c}$ approximations in the framework of the formalism, see \cite{Our1, Hetch}, allowed to establish a connection between the
Lipatov's approach to the high energy processes and the CGC (Color Glass Condensate), BK and Balitsky-JIMWLK approaches to high energy scattering
\cite{Venug,BK,Kovner,Hatta1}.
Besides these approximations, the another difference between the approaches is the degrees of freedom under consideration, as it was noted above.
In the effective action formalism these are reggeized gluons fields, see
\cite{BFKL,LipatovEff}, whereas the shock wave formalisms operates with the correlators of Wilson lines constructed from the gluon or quark fields. These degrees of freedom are
different but are complimentary each to other in the sense that the Lipatov's effective action can be considered as reformulation of the theory of interacting  operators\footnote{
We will call the operators based on the combinations of Wilson lines as Lipatov's operators further as well.} of Wilson lines
without any Reggeons introduced, see details and discussion in \cite{Our4}. Therefore, the new formulation of the effective action approach proposed in \cite{Our4} allows to establish the
connection between the correlators of the reggeized gluon fields  and correlators of Lipatov's operators, i.e. it allows to establish the connection between two main but different degrees of freedom
in high energy QCD beyond any approximations.

  The equations
for the correlators of the reggeized fields in the QCD RFT must be derived in the Lipatov's effective action formalism in the form of BFKL like equations for any correlators of interests.
The straightforward way to achieve that is a derivation of the Dyson-Schwinger hierarchy of the equations for the correlators starting directly from the generating functional
for the reggeized fields, this is the main goal of the present  note and it is considered in the second Section of the paper. The perturbative expressions for the  Reggeon fields correlators  till correlator of four-fields inclusively
are written in Appendixes A-D, where we solved perturbatively the hierarchy of the correlator's equations and write explicitly the form of the corrections to the
BFKL like  correlators of one, two, three and four Reggeons.  In Section 3 we discuss the leading order BK like equation for the correlator of two Reggeon fields which form
depends on the perturbative scheme used in the calculations. The connection between the correlators of
Lipatov's operators and reggeized fields correlators is established and discussed in Section 4, where also the precise equation for the connection of the
correlator of two Lipatov's operators with correlator of two Reggeon fields is written. The last Section 5 is the conclusion of the paper.

\section{Lipatov's effective action and correlators of Reggeon fields }

 The Lipatov's effective action for reggeized gluons $\B_{\pm}$, formulated as  RFT (Regge Field Theory), can be obtained by an integration out the gluon fields $\tv$ in the
generating functional for the $S_{eff}[\tv,\,\B]$:
\beq\label{Add1}
e^{\imath\,\Gamma[\B]}\,=\,\int\,D \tv\,e^{\imath\,S_{eff}[\tv,\,\B\,]}
\eeq
where
\beq\label{Add2}
S_{eff}\,=\,-\,\int\,d^{4}\,x\,\Le\,\frac{1}{4}\,G_{\mu \nu}^{a}\,G^{\mu \nu}_{a}\, \,+\,tr\,\left[\,\Le\,\T_{+}(\textsl{v}_{+})\,-\,\B_{+}\,\Ra\,j_{reg}^{+}\,+\,
\Le\,\T_{-}(\textsl{v}_{-})\,-\,\B_{-}\,\Ra\,j_{reg}^{-}\,\right]\,\Ra\,,
\eeq
with
\beq\label{Add3}
\T_{\pm}(\textsl{v}_{\pm})\,=\,\frac{1}{g}\,\D_{\pm}\,O(\textsl{v}_{\pm})\,=\,
\textsl{v}_{\pm}\,O(\textsl{v}_{\pm})\,,\,\,\,\,
j_{reg\,a}^{\pm}\,=\,\frac{1}{C(R)}\,\D_{i}^{2}\,\B_{a}^{\pm}\,,
\eeq
here  $C(R)$ is eigenvalue of Casimir operator in the representation R, $tr(T^{a} T^{b})\,=\,C(R)\,\delta^{a b}$\, see \cite{LipatovEff,Our1,Our2,Our3}.
The form of the Lipatov's operator  $O$ (and correspondingly $\T$) depends on the particular process of interests,
in the non-Hermitian case it has the form of  the Wilson line (ordered
exponential) for the longitudinal gluon fields in an arbitrary representation:
\beq\label{Add4}
O(\tv_{\pm})\,=\,P\,e^{g\,\int_{-\infty}^{x^{\pm}}\,d x^{\pm}\,\tv_{\pm}(x^{+},\,x^{-},\,x_{\bot})}\,,\,\,\,\,\tv_{\pm}\,=\,\imath\,T^{a}\,\tv_{\pm}^{a}
\eeq
see details in \cite{LipatovEff, Our4}. A Hermitian  form of the operators was also derived in \cite{Our4}, in this case the operator is
represented by a combination of the different Wilson lines:
\beqar\label{Add41}
O(\tv_{\pm})\,& = &\frac{1}{4}
\Bigl(
P\,e^{g\,\int_{-\infty}^{x^{+}}\,d x^{\pm}\, \tv_{\pm}(x^{+},\,x^{-},\,x_{\bot})}-
P\,e^{g\,\int_{x^{+}}^{\infty}\,d x^\pm\,\tv_{\pm}(x^{+},\,x^{-},\,x_{\bot})}-\nonumber \\
&-&\,\bar{P}\,e^{- g \,\int_{-\infty}^{x^{+}}\,d x^{\pm}\, \tv_{\pm}(x^{+},\,x^{-},\,x_{\bot})}\,+
\bar{P}\,e^{-g \,\int_{x^{+}}^{\infty}\,d x^{\pm}\, \tv_{\pm}(x^{+},\,x^{-},\,x_{\bot})}\,
\Bigr)\,,
\eeqar
see also \cite{Nefedov}.

 The effective action $\Gamma$ of the interactions of  reggeized gluons was calculated to one-loop precision in \cite{Our2} for the case of the adjoint representation of gluon fields.
This actions has the following form\footnote{In order to make the notations shorter further we do not write explicitly the coordinate dependence of the functions
and use the following expressions for the vertices:
for the vertices, i.e. we write:
$K^{+\,a_1,\,\cdots\,,+\,a_n; -\,b_1,\,\cdots\,,-\,b_m}\,=\,
\Le K^{a_1\,\cdots\,a_n}_{b_1\cdots\,b_m} \Ra^{+\,\cdots\,+}_{-\,\cdots\,-\,}\,$.} :
\beq\label{Sec1}
\Gamma\, = \,\sum_{n,m\,=\,1}\,\Le\,\B_{+}^{\,a_1}\,\cdots\,\B_{+}^{\,a_n}\,\Le K^{+\,\cdots\,+}_{-\,\cdots\,-}\Ra^{a_1\cdots\, a_n}_{b_1\cdots\, b_m}\,\B_{-}^{\,b_1}\,\cdots\,\B_{-}^{\,b_m}\,\Ra\,=\,
\,-\,\B_{+\,x}^{\,a}\,\D_{i}^{2}\,\B_{-\,x}^{\,a}\,+\,\B_{+\,x}^{\,a}\,\Le K_{x y}^{\,a\, b} \Ra^{+}_{-}\,\B_{-\,y}^{\,b}\,+\,\cdots\,,
\eeq
where $\B_{\pm}$ are the reggeized gluon fields and shorthand notations for the integration over the variables in the action were used.
The effective vertices  (kernels\footnote{In the language of high-energy perturbative QCD these vertices are BFKL-like  kernels of the integro-differential equations for the
 objects of interests or, equivalently, they can be considered as analog of the different parts of a Hamiltonian in Balitsky-JIMWLK approach, see in \cite{Venug,BK,Kovner,Hatta1}.})
$K$ in \eq{Sec1} represents
the processes of  multi-Reggeon interaction in t-channel of the high energy scattering amplitude.
For example, the vertex responsible for the gluon's reggeization has the following form\footnote{We did not present here the parts of the vertex arising as pure QCD corrections
to the propagator, see \cite{Our2}. These corrections provide sub-leading contributions in the Regge kinematics of the formalism. Nevertheless, the expressions of Appendixes A-D are correct in the general case as well
if in \eq{B4} equation we will understand $K^{+}_{-}$ vertex as the full one.} to LO:
\beqar\label{Sec2}
\Le\,K^{\,a\,b}_{x\,y}\,\Ra_{-\,}^{+}\,& = & \, \Le\,K^{\,a}_{\,b}\,\Ra_{-\,}^{+}\, =\, -\delta(x^{+}-y^{+})\,\delta(x^{-}-y^{-})\,\delta^{\,a\,b}\nonumber \\
 &\,&\frac{g^{2} N}{8 \pi}\, \D_{i\,x}^{2}
\Big(\int_{\Lambda}^{\Lambda\,e^{\eta}}\frac{d p_{-}}{p_{-}}
\int \frac{d^{2} p_{\bot}}{(2 \pi)^{2}}
\int \frac{d^{2} k_{\bot}}{(2 \pi)^{2}}
\frac{k_{\bot}^{2}}{p_{\bot}^{2} \Le\,p_{\bot}-k_{\bot}\Ra^{2}}e^{-\imath\,k_{i} \Le x_{i}-y_{i}\Ra}\Big),
\eeqar
see details of the calculations in \cite{Our2}. The rapidity interval $\eta$
in \eq{Sec2} is an analog of the ultraviolet cut-off in the relative longitudinal momenta.
Physically it determines the value of the cluster of the particles in the Lipatov's effective action approach, see \cite{LipatovEff}.
In order to calculate the correlators of the Reggeon $\B_{\pm}$ fields, we will use the following generating functional for Reggeon fields:
\beq\label{Sec21}
Z[J]\, = \,\int\,D\B\,{\rm exp}\,\Big(\imath\,\Gamma[\B]-\imath\int\,d^4x\,J_{-}^{\,a}(x^{-},\,x_{\bot})\,\B_{+}^{\,a}(x^{+},\,x_{\bot})\,-\,
\imath\int\,d^4x\,J_{+}^{\,a}(x^{+},\,x_{\bot})\,\B_{-}^{\,a}(x^{-},\,x_{\bot})\,\Big)\,,
\eeq
see \cite{Our4}, with some auxiliary currents $J_{\pm}$ introduced.
Correspondingly,
the Schwinger-Dyson equations for the correlators we obtain now taking derivative of the field's variation of $Z[J]$ in respect to the currents
and taking them equal to zero at the end.
Namely, to the first order we have:
\beq\label{Sec24}
\delta\,Z[J]= \int D\B\,\delta \B_{\pm}^{a}\Le\frac{\delta \Gamma [\B]}{\delta \B_{\pm}^{\,a}}-\int\,dx^{\mp}\,J_{\mp}^{\,a}(x^{\mp},\,x_{\bot})\Ra
{\rm exp}\Big({\imath \Gamma[\B]-\imath\,\int\,d^4x\,J_{-}^{\,a}\,\B_{+}^{\,a}
-\imath \int\,d^4x\,J_{+}^{\,a}\,\B_{-}^{\,a}}\Big) =0\,.
\eeq
Therefore, for the one-field correlator we obtain:
\beq\label{Sec3}
<\,\frac{\delta \Gamma [\B]}{\delta \B_{\pm}^{\,a}}\,>\,=\,0\,
\eeq
that, using  \eq{Sec1} expansion, provides\footnote{Here the coordinate dependence is written precisely and averaging means, as usual, the averaging
of the fields with the use of \eq{Sec21} functional taking $J\,=\,0$ limit after the calculations.
The summation in the expression over the $B^{a_i}_{\pm}$ Reggeon fields must be understood in the sense that
 for each given $i$ the
both $+$ and $-$ signs and all their combinations are accounted in the sum, i.e. for the $n\,=\,1$, for example, there are two terms in the
r.h.s. of  \eq{Sec4} which correspond to the $B^{a_1}_{\,+}$ and $B^{a_1}_{-}$ Reggeon fields in the corresponding terms, see Appendixies A-D.}:
\beq\label{Sec4}
<\,\B_{\pm}^{\,a}(x,\,\eta)\,>\,=\,
\sum_{n\,=\,1}\,\Le\,\tilde{K}(\eta; x,\,x_{1}\,\cdots,\,x_{n})\,\Ra^{\,a\,a_{1}\,\cdots\,a_{n}}\,
<\,\B_{\pm\,}^{\,a_1}(x_{1})\,\cdots\,\B_{\pm\,}^{\,a_n}(x_{n})\,>\,,
\eeq
see Appendix A. In the r.h.s. of the expression we introduce the shorthand notation $\tilde{K}$ for the redefined vertices
obtained by the convolution of \eq{Sec1} vertices and bare propagators of the theory, see Appendixes A-D.
Taking the derivative from the \eq{Sec4} with respect to the $\eta$, we can also obtain the BFKL like evolution equation for the reggeized fields:
\beq\label{Sec5}
\frac{\D\,}{\D\,\eta}\,<\,\B_{\pm}^{\,a}\,>\,=\,\frac{\D}{\D\,\eta}\,\Le\,
\sum_{n\,=\,1}\,\Le\,\tilde{K}(\eta)\,\Ra^{\,a\,a_{1}\,\cdots\,a_{a}}\,<\,\B_{\pm\,}^{\,a_1}\,\cdots\,\B_{\pm\,}^{\,a_n}\,>\,
\Ra\,,
\eeq
we see from the equation, that indeed, after the derivation in respect to $\eta$ in the r.h.s. of the equation, the effective vertices
will form corresponding parts of the Hamiltonian responsible for the evolution of the fields in the rapidity space.
The next derivative of \eq{Sec24} with respect to the currents provides the following equation\footnote{Further we will
use $\delta^{a b}$ note for the notation of 4-delta function plus color indexes included, denoting full coordinate and color indexes dependence in the delta functions
only where it will be needed.}:
\beq\label{Sec31}
<\,\frac{\delta \Gamma [\B]}{\delta \B_{\pm\,}^{\,a_1}}\,\B_{\pm\,}^{\,a_2}\,-\,\imath\,\delta^{\,a_1\,a_2}\,\delta_{\pm 1\,\,\pm 2}\,\delta (x_{\bot\,1}\,-\,x_{\bot\,2})\,>\,=\,0\,.
\eeq
Considering the expression for the $<\,\B_{\pm}\,\B_{\mp}\,>$ correlation function in adjoint representation, we correspondingly we obtain:
\beq\label{Sec32}
\D_{\bot\,1}^{2}\,<\,\B_{\pm\,}^{\,a_1}\,\B_{\mp\,}^{\,a_2}\,>\,=\,-\,\imath\,\delta^{\,a_1\,a_2}\,\delta_{\pm 1\,\,\pm 2}\,\delta (x_{\bot\,1}\,-\,x_{\bot\,2})\,+\,
\Le\,K^{\,a_1\,}_{\,b_{1}\,}\,\Ra^{\pm\,}_{\,\mp}\,
<\B_{\pm\,}^{\,b_1}\,\B_{\mp\,}^{\,a_2}>\,+\,\cdots\,.
\eeq
We see here that this equation indeed reproduces the equation for the propagator of the reggeized fields, see \cite{Our2} and \eq{B4}.

 The general Schwinger-Dyson system of the equations for the field's correlators, correspondingly, can be written as
\beq\label{Sec33}
<\,\frac{\delta \Gamma [\B]}{\delta \B_{\pm}^{\,a}}\,\B_{\pm}^{\,a_1}\cdots \B_{\pm}^{\,a_{n}}\,>_{J}\,-\,\imath\,\sum_{i=1}^{n}\,<\,
\B_{\pm}^{\,a_{1}}\cdots \delta^{\,a\,a_i}\,\delta_{\pm\,a \,\pm\,a_{i}}\,\delta (x_{\bot\,a}\,-\,x_{\bot\,a_{i}})\cdots \B_{\pm}^{\,a_{n}}\,>_{J}\,=\,0\,,
\eeq
that, with the help of \eq{Sec1}, provides
\beq\label{Sec6}
<\,\B_{\pm\,}^{\,a}\,\B_{\pm\,}^{\,a_{1}}\,\cdots\,\B_{\pm\,}^{\,a_{m}}\,>\,=\,
\sum_{n\,=\,1}\,\Le\,\hat{K}(\eta)\,\Ra^{\,a\,b_{1}\cdots b_{n}}
<\,\B_{\pm\,}^{\,b_{1}}\,\cdots\,\B_{\pm\,}^{\,b_{n}}\,\B_{\pm\,}^{\,a_{1}}\,\cdots\,\B_{\pm\,}^{\,a_{m}}\,>\,.
\eeq
Examples of the equations for the two, three and four Reggeon fields correlators are presented in the Appendixes.
Now, taking dervatives of the l.h.s. and r.h.s. of the \eq{Sec6} with respect to the parameter $\eta$ we obtain the same system
of the	equations as a system of BFKL like coupled evolution equations for the correlators:
\beq\label{Sec7}
\frac{\D}{\D\, \eta}<\,\B_{\pm\,}^{\,a}\,\B_{\pm\,}^{\,a_{1}}\,\cdots\,\B_{\pm\,}^{\,a_{m}}\,>\,=\,\frac{\delta}{\delta\, \eta}\,
\sum_{n\,=\,1}\,\Le\,\hat{K}(\eta)\,\Ra^{\,a\,b_{1}\cdots b_{n}}
<\,\B_{\pm\,}^{\,b_{1}}\,\cdots\,\B_{\pm\,}^{\,b_{n}}\,\B_{\pm\,}^{\,a_{1}}\,\cdots\,\B_{\pm\,}^{\,a_{m}}\,>\,.
\eeq	
Formally, the form of \eq{Sec7} is similar to the Balitsky-JIMWLK hierarchy of the correlators of Wilson lines. In the paper \cite{Our4} the connections between the correlators
of the Reggeon fields and operators constructed from the Wilson lines was established, therefore further we will use the results of \cite{Our4} in order to derive  hierarchy
of Lipatov's operators in terms of Reggeon corelators.

 The structure of the rapidity dependence in \eq{Sec4}-\eq{Sec5} or \eq{Sec32}-\eq{Sec7} can be clarified on the base of BFKL calculus, \cite{BFKL}, see also \cite{Our1,Our2}
for the calculations made in the given framework. Namely, taking
Fourier transform of the correlators it is assumed that any correlator of interests, to the LO values of the effective vertices, can be written
similarly to the equation for the propagator of reggeized gluons, see \cite{Our2}, i.e.
as some
integral equation\footnote{We do not write here precisely the momenta dependence and integration on the momenta
of the correlators and vertices in the equation, see some calculations in \cite{Our2}.} of the following form :
\beq\label{Sec71}
<\,\tilde{\B}_{\pm\,}^{\,a}\,\tilde{\B}_{\pm\,}^{\,a_{1}}\,\cdots\,\tilde{\B}_{\pm\,}^{\,a_{m}}\,>_{\eta}\,=\,
\sum_{n\,=\,1}\,\int_{0}^{\eta}\,d \eta^{'}\Le\,\hat{\tilde{K}}(\eta^{'})\,\Ra^{\,a\,b_{1}\cdots b_{n}}
<\,\tilde{\B}_{\pm\,}^{\,b_{1}}\,\cdots\,\tilde{\B}_{\pm\,}^{\,b_{n}}\,\tilde{\B}_{\pm\,}^{\,a_{1}}\,\cdots\,\tilde{\B}_{\pm\,}^{\,a_{m}}\,>_{\eta^{'}}\,,
\eeq
where the correlators and vertices are depend only on corresponding two-dimensional transverse momenta, see \eq{Sec2} and \cite{Our2} for more details.
Taking the derivative on rapidity from the both sides of the equation, accordingly to \eq{Sec7},
we will obtain the BFKL like equations for the correlators, this is the way how the BFKL calculus is arising in the formalism of Lipatov's effective action. Considering for example the propagator
of two reggeized gluons in momentum space, i.e. $<\,\tilde{\B}_{\,+}^{\,a}\,\tilde{\B}_{\,-}^{\,b}\,>\,$ correlator, see \cite{Our2},
we immediately will conclude that the reggeization of the propagator  means the
preservation of the form of the \eq{Sec71} with corresponding momenta dependence to the all perturbation orders of the theory with all effective vertices accounted which is not the case in general, see \cite{FadLipLast} for example.

\section{Leading order solutions for two-fields correlator}

 The BFKL like equation for the two-fields correlator, \eq{B3}, can be written with the help of Appendixes A-D results.
The solution of the equation can be obtained or by use of the perturbative expansion \eq{B5}
or by writing and subsequent solution of the corresponding BFKL type equation
with non-linear corrections accounted. Further, in both case, we consider only
the leading in the BFKL sense contributions, see Appendix A, and note that
the form of the equation depends in general on the perturbative scheme of the calculations chosen.

  First of all, consider the perturbative solution of the \eq{B3} equation.
In the RFT scheme the bare propagator
of the scheme is given by \eq{B4} and instead of an analog of JIMWLK-Balitsky equation
\beq\label{E8}
\D^{2}_{\bot}\,<\B_{+}^{\,a}\,\B_{-}^{\,a_1}>\,=\,-\,\imath\,\delta^{\,a a_1}\,+\,\Le K_{\,a}^{\,a_2} \Ra^{+}_{-}\,<\B_{+}^{\,a_2}\,\B_{-}^{\,a_1}>\,+\,
\,2\,\Le K_{a\,a_2}^{\,a_3\,a_4}\Ra_{- - }^{+ +}\,<\B_{+}^{\,a_3}\,\B_{+}^{a_4}\,\B_{-}^{a_2}\,\B_{-}^{\,a_1}>\,
\eeq
the following expression will be obtained
\beqar\label{E81}
\D^{2}_{\bot}\,<\B_{+}^{\,a}\,\B_{-}^{\,a_1}>\,& = &\,-\,\imath\,\delta^{\,a a_1}\,+\,\Le K_{\,a}^{\,a_2} \Ra^{+}_{-}\,<\B_{+}^{\,a_2}\,\B_{-}^{\,a_1}>\,-\,
\,2\,\imath\,\Le K_{a\,a_2}^{\,a_3\,a_4}\Ra_{- - }^{+ +}\,<\B_{+}^{\,a_3}\,\B_{-}^{a_2}>\,G_{0\,-}^{\,a_4 a_1\,+}\,-\,\nonumber \\
&-&\,2\,\imath\,\Le K_{a\,a_2}^{\,a_3\,a_4}\Ra_{- - }^{+ +}\,G_{0\,-}^{\,a_3 a_2\,+}\,<\B_{+}^{\,a_4}\,\B_{-}^{a_1}>\,
\eeqar
after insertion of \eq{D3} in \eq{B3}.
Similarly to routines of the BFKL calculus,
we assume\footnote{The $K^{++}_{--}$ vertex is known from the BFKL physics but  did not calculated in the framework of \cite{Our1,Our2} formalism yet.} that
the vertex
$K^{++}_{--}$ is local in rapidity space. Therefore we note, that to the given precision
we can rewrite  the equation as
\beq\label{E811}
\D^{2}_{\bot}\,<\B_{+}^{\,a}\,\B_{-}^{\,a_1}>\ = \,-\,\imath\,\delta^{\,a a_1}\,+\,\Le K_{\,a}^{\,a_2} \Ra^{+}_{-}\,<\B_{+}^{\,a_2}\,\B_{-}^{\,a_1}>\,-\,
4\,\imath\,\Le K_{a\,a_2}^{\,a_3\,a_4}\Ra_{- - }^{+ +}\,G_{0\,-}^{\,a_3 a_2\,+}\,<\B_{+}^{\,a_4}\,\B_{-}^{a_1}>\,.
\eeq
Namely, the precision of the \eq{E81}  is defining correspondingly to the precision determined by the
$<\B_{+}^{\,a_3}\,\B_{-}^{a_2}>\,\rightarrow\,-\,\imath\,G_{0\,-}^{\,a_3 a_2\,+}\,$ substitution in the
nonlinear second term of \eq{E81} equation. Therefore, introducing
a redefined  effective vertex  in the equation
\beq\label{E1}
\Le \hat{K}_{b\,}^{\,b_3}\Ra_{-  }^{+}\,=\,\imath\,\Le K_{b\,b_1}^{\,b_2\,b_3}\Ra_{- - }^{+ +}\,G_{0\,-}^{\,b_2 b_1\,+}\,=
\,-\,\Le K_{b\,b_1}^{\,b_2\,b_3}\Ra_{- - }^{+ +}\,<\B_{+}^{\,b_2}\,\B_{-}^{\,b_1}>_{0}\,,
\eeq
and assuming that the new redefined vertex is preserving the "reggeization" property of the
approach, see \eq{Sec71} above, we arrive to the following equation:
\beq\label{E12}
\Le\D^{2}_{\bot}\,\delta^{a b}-\Le K_{\,a}^{\,b} \Ra^{+}_{-} + 4\,\Le \hat{K}_{\,a}^{\,b}\Ra_{-  }^{+ }\,\Ra\,
<\B_{+}^{\,b}\,\B_{-}^{\,d}>\,=\,-\,\imath\,\delta^{\,a d}
\eeq
that corresponds to the redefinition of the trajectory of the reggeized gluons propagator as
\beq\label{E4}
K^{+}_{-}\,\rightarrow\,K^{+}_{-}\,-\,4\,\Le \hat{K}\Ra_{- }^{+}\,,
\eeq
which correspond to additional one-loop RFT tadpole contribution to the trajectory, see \eq{E811} and also calculations in \cite{Our2}.
If the "reggeization" property does not hold for the \eq{E1} vertex then
using \eq{B5} perturbative expansion we obtain to the required perturbative order:
\beq\label{E2}
<\B_{+}^{\,a}\,\B_{-}^{\,a_1}>_1\,=\,\,4\,\imath\,G_{0\,-}^{\,a b\,+}\,\Le \hat{K}_{b\,}^{\,b_1\,}\Ra_{- }^{+}\,G_{0\,-}^{\,b_1 a_1\,+}\,,
\eeq
that is the same correction which can be obtained directly from expansion of \eq{E12}, but which holds only to particular perturbative order in contrast to \eq{E4} expression.

 The equation for the two-fields correlator can be written differently if instead the perturbative solution we consider a BFKL like equation for the correlator.
In this case, the \eq{D1} can be written with the prescribed precision as
\beq\label{E13}
<\B_{+}^{\,a_3}\,\B_{+}^{a_4}\,\B_{-}^{a_2}\,\B_{-}^{\,a_1}>\,=\,2\,<\B_{+}^{\,a_3} \B_{-}^{\,a_2}>\,<\B_{+}^{\,a_4} \B_{-}^{\,a_1}>\,.
\eeq
Indeed, inserting this factorized expression back in the \eq{D1}, we will obtain the following expression for the two fields correlator to
leading order precision:
\beq\label{E14}
\D^{2}_{\bot}\,<\B_{+}^{\,a}\,\B_{-}^{\,a_1}>\,=\,-\,\imath\,\delta^{\,a a_1}\,+\,\Le K_{\,a}^{\,a_2} \Ra^{+}_{-}\,<\B_{+}^{\,a_2}\,\B_{-}^{\,a_1}>\,,
\eeq
which is the same as \eq{B3} that justifies \eq{E13} assumption. Therefore, to leading order, we have for \eq{B3}:
\beq\label{E15}
\D^{2}_{\bot}\,<\B_{+}^{\,a}\,\B_{-}^{\,a_1}>\,=\,-\,\imath\,\delta^{\,a a_1}\,+\,\Le K_{\,a}^{\,a_2} \Ra^{+}_{-}\,<\B_{+}^{\,a_2}\,\B_{-}^{\,a_1}>\,+\,
\,4\,\Le K_{a\,a_2}^{\,a_3\,a_4}\Ra_{- - }^{+ +}\,<\B_{+}^{\,a_3} \B_{-}^{a_2}>\,< \B_{+}^{a_4}\,\B_{-}^{\,a_1}>\,.
\eeq
If in this expression the tadpole-like term we will replace the full correlator with the bare one, similarly to done above, we will arrive to \eq{E12} again.
Without this replacement, introducing the
usual BFKL bare propagator
\beq\label{E6}
\delta^{a\,b}\,\delta^{4}(x\,-\,y)\, \D^{2}_{\bot\,x}\,\hat{G}_{0\,-}^{\,b c\,+}(y,z)\,=\,\delta^{a\,c}\,\delta^{4}(x\,-\,z)\,
\eeq
and
preserving the BFKL like form of the equation, we will obtain:
\beqar\label{E16}
<\B_{+}^{\,a}\,\B_{-}^{\,a_1}>\,&=&\,-\,\imath\,\hat{G}_{0\,-}^{\,a a_1\,+}\,+\,\hat{G}_{0\,-}^{\,a b\,+}\Le K_{\,b}^{\,b_1} \Ra^{+}_{-}\,<\B_{+}^{\,b_1}\,\B_{-}^{\,a_1}>\,+\,
\nonumber \\
&+&\,4\, \hat{G}_{0\,-}^{\,a b\,+}\Le K_{b\,b_1}^{\,b_2\,b_3}\Ra_{- - }^{+ +}\,<\B_{+}^{\,b_2} \B_{-}^{b_1}>\,< \B_{+}^{b_3}\,\B_{-}^{\,a_1}>\,.
\eeqar
This equation we can consider as analog of BK equation written for the correlator of the reggeized gluon fields in the framework with \eq{E6} propagator use.
In general, the form of the solution of this equation depends on the bare propagator, i.e. in the framework of RFT formulation with \eq{B4} propagator, the
equation \eq{E16} will achieve the following form
\beq\label{E161}
<\B_{+}^{\,a}\,\B_{-}^{\,a_1}>\,=\,-\,\imath\,G_{0\,-}^{\,a a_1\,+}\,+\,
4\,G_{0\,-}^{\,a b\,+}\Le K_{b\,b_1}^{\,b_2\,b_3}\Ra_{- - }^{+ +}\,<\B_{+}^{\,b_2} \B_{-}^{b_1}>\,< \B_{+}^{b_3}\,\B_{-}^{\,a_1}>\,
\eeq
which solution will depend on the rapidity ordering of the functions in the second term in the r.h.s. of the expression, see \eq{Sec71}.
Can the \eq{E16} and \eq{E161} be solved similarly to the ways of solution of usual BFKL equation is not clear, we plane to investigate this problem in some separate publication.

\section{Correlators of Wilson lines operators }

 In this Section we consider the generating functional for the Lipatov's operators introduced above in \eq{Add3}:
\beqar\label{WL2}
Z[J] \,&= &\,
\frac{1}{Z^\prime} \int D\tv \, {\rm exp}\,\Big(\imath\, S^{0}[\tv]\,-\,
\frac{\imath}{2\,C(R)}\int\, d^4 x \,\T_{+}\,\partial_{\bot}^{2}\,\T_{-}\,-\,\nonumber \\
&-&\,\frac{\imath}{2\,C(R)}\int\, d^4 x \,J_{-}(x^{-},\,x_{\bot})\,\T_{+}\,-\,
\frac{\imath}{2\,C(R)}\int\, d^4 x \,J_{+}(x^{+},\,x_{\bot})\,\T_{-}\,\Big)\,,
\eeqar
with $S^{0}$ as the gluon's action on which the averaging of the operators is performed\footnote{For shortness we use the same notation for the generating functional as in  \eq{Sec21}.}.
Taking, for example, two derivatives of $\log Z[J]$ with respect to the currents we obtain:
\beqar\label{WL21}
&\,&\,C(R)^2\,(\,2\,\g)^{2}\,\Le\,\frac{\delta^{2}}{\delta J_{1\,\mp}^{a_1}\,\delta J_{2\,\mp}^{a_2}}\,\log Z[J]\,\Ra_{\,J\,=\,0}\, = \,\nonumber \\
&=&\,<\,T^{a_1}\,\Le O_{1}(\tv_{\pm})_{x^{\pm}\,=\,\infty}\,-\,O_{1}(\tv_{\pm})_{x^{\pm}\,=\,-\infty}\, \Ra\,\otimes\,T^{a_2}
\Le O_{2}(\tv_{\pm})_{x^{\pm}\,=\,\infty}\,-\,O_{2}(\tv_{\pm})_{x^{\pm}\,=\,-\infty}\, \Ra\,>\,=\, \nonumber \\
&=&\,<\,\Le T^{a_1}\,\hat{O}_{1}(\tv_{\pm})\Ra\,\otimes\,\Le T^{a_2}\,\hat{O}_{2}(\tv_{\pm})\Ra\,>\,,
\eeqar
that in the case of \eq{Add4} representation, for example, provides us with the correlators of usual Wilson lines:
\beq\label{WL1}
\hat{O}(\tv_{\pm})\,=\,W_{\pm\,}\,=\,P\,e^{g\,\int_{-\infty}^{\infty}\, d x^{\pm}\,\tv_{\pm}(x^{\pm},\,x^{\mp},\,x_{\bot})}\,-\,1\,,\,\,\,\,\tv_{\pm}\,=\,\imath\,T^{a}\,A_{\pm}^{a}\,,
\eeq
whereas in the case of \eq{Add41} representation of the operators the correlators of the following operators arise:
\beq\label{WL11}
\hat{O}(\tv_{\pm})\,=\,\frac{1}{2}\,\Le\,
P\,e^{g\,\int_{-\infty}^{\infty}\, d x^{\pm}\,\tv_{\pm}(x^{\pm},\,x^{\mp},\,x_{\bot})}\,-\,
\bar{P}\,e^{-\,g\,\int_{-\infty}^{\infty}\, d x^{\pm}\,\tv_{\pm}(x^{\pm},\,x^{\mp},\,x_{\bot})}\,\Ra\,,
\eeq
these operators we also can call as Lipatov's operators.
In accordance with \cite{Our4}, we can rewrite the generating functional \eq{WL2}  with the help of new degrees of freedom, which we identify as the reggeized gluons fields:
\beqar\label{WL4}
Z[J] \,&= &\,
\frac{1}{Z^\prime} \int D\tv \, {\rm exp}\,\Big(\imath\, S^{0}[\tv]\,-\,\frac{\imath}{2\,C(R)}\,\int d^4 x\,
\Le -\D^{2}_{\bot}\T_{+} - J_{+}  \Ra\,\Le \D^{2}_{\bot} \Ra^{-1}\,\Le -\D^{2}_{\bot}\T_{-} - J_{-}  \Ra\,+\,\nonumber \\
&+&\,\frac{\imath}{2\,C(R)} \int d^{4}x \,J_{+}(x^{+},\,x_{\bot})\,\Le\D_{\bot}^{2}\Ra^{-1} J_{-}(x^{-},\,x_{\bot})\,\Big)\,=\, \nonumber \\
&=&\,
\frac{1}{Z^\prime} \int D\tv \, D\B\, {\rm exp}\,\Big(\imath\, S^{0}[\tv]\,+\,
\frac{2\,\imath}{\,C(R)}\,\int d^4 x \,\B_{+}(x^{+},\,x_{\bot})\,\partial_{\bot}^{2}\,\B_{-}(x^{-},\,x_{\bot})\,-\,\nonumber \\
&-&\,\frac{\imath}{\,C(R)}\,\int\, d^4 x \,\,\T_{+}(x^{+},\,x^{-},\,x_{\bot})\,\D^{2}_{\bot}\,\B_{-}(x^{-},\,x_{\bot})\,-\,
\frac{\imath}{\,C(R)}\,\int\, d^4 x \,\,\T_{-}(x^{+},\,x^{-},\,x_{\bot})\,\D^{2}_{\bot}\,\B_{+}(x^{+},\,x_{\bot})\,-\,\nonumber \\
&-&\,\frac{\imath}{\,C(R)}\,\int\, d^4 x \,\,J_{-}(x^{-},\,x_{\bot})\,\B_{+}(x^{+},\,x_{\bot})\,-\,
\frac{\imath}{\,C(R)}\,\int\, d^4 x \,\,J_{+}(x^{+},\,x_{\bot})\,\B_{-}(x^{-},\,x_{\bot})\,+\,\nonumber \\
&+&\,\frac{\imath}{2\,C(R)} \int d^{4}x \,J_{+}(x^{+},\,x_{\bot}) \Le\D_{\bot}^{2}\Ra^{-1} J_{-}(x^{-},\,x_{\bot})\,\Big)\,
\eeqar
and which also can be written as
\beqar\label{WL41}
Z[J] \,&=&\,
\frac{1}{Z^\prime} \int D\tv \, D\B\, {\rm exp}\,\Big(\imath\, S_{eff}[\tv,\,\B\,]\,+\,
\frac{\imath}{2} \int d^{4}x \,J_{+}^{\,a}(x^{+},\,x_{\bot}) \Le\D_{\bot}^{2}\Ra^{-1} J_{-}^{\,a}(x^{-},\,x_{\bot})\,-\,\nonumber \\
&-&\,\imath\,\int\, d^4 x \,\,J_{+}^{\,a}(x^{+},\,x_{\bot})\,\B_{-}^{\,a}(x^{-},\,x_{\bot})\,-\,
\imath\,\int\, d^4 x \,\,J_{-}^{\,a}(x^{-},\,x_{\bot})\,\B_{+}^{\,a}(x^{+},\,x_{\bot})\,\Big)\,,
\eeqar
see \eq{Add2} and \eq{Sec21} definitions.
Now we can define an arbitrary correlator of the $\hat{O}$ operators as
\beqar\label{Sec12}
&&<\,\Le\,T^{a_1}\,\hat{O}_{1}(\tv_{\pm})\Ra\,\otimes\,\Le\,T^{a_2}\,\hat{O}_{2}(\tv_{\pm})\Ra\,\otimes\,\cdots\,\otimes\,
\Le\,T^{a_n}\,\hat{O}_{n}(\tv_{\pm})\Ra\,>\,= \,\nonumber \\
&=&\,C(R)^{n}\,(\,2\,g)^{n}\,
\Le\,\frac{\delta^{\,n}}{\delta J^{a_1}_{\mp\,}\,\cdots\,\delta J^{a_n}_{\mp\,}}\,\log Z[J] \Ra_{J\,=\,0\,}\,.
\eeqar
The general expression for the r.h.s of \eq{Sec12} is cumbersome\footnote{The structure of the answer is easy reconstructed if
we put attention that a relative dimension of the $\D^{-2}_{\bot}$ operator in terms of $B$ or $\hat{O}$ operators is 2.Therefore, requesting that the dimension of the
r.h.s and l.h.s of the \eq{Sec12} will be equal we arrive to the following expression:
$$
<\,\hat{O}_{1}\,\cdots\,\hat{O}_{n}>\,\propto\,\sum_{j\,=\,0}\,\D^{-2j}_{\bot}\,<\,\B_{1}\,\cdots\,\B_{n-2j}>\,,
$$
where the sum over $j$ is going till $n/2$ or $(n-1)/2$ in the case of even or odd $n$ correspondingly.}
therefore we write the expression in the following form:
\beqar\label{Sec121}
&&<\,\Le\,T^{a_1}\,\hat{O}_{1}(\tv_{\pm})\Ra\,\otimes\,\Le\,T^{a_2}\,\hat{O}_{2}(\tv_{\pm})\Ra\,\otimes\,\cdots\,\otimes\,
\Le\,T^{a_n}\,\hat{O}_{n}(\tv_{\pm})\Ra\,>\,= \,\nonumber \\
&=&C(R)^{n} \Le 2 g \Ra^{n} \int dx_{1}^{\pm}\cdots\int dx_{\,n}^{\pm}\Big(
\sum_{j=0}^{J} C_{b_1\cdots\,b_{n-2j}}^{\,a_1 \cdots a_{n}} \Le \frac{\imath}{2}\,\D^{-2}_{\bot} \Ra^{j} \Le -\imath \Ra^{n-2j}
<\B_{\pm}^{\,b_1}\,\cdots\,\B_{\pm}^{\,b_{n-2j}}>\Big),
\eeqar
where $J\,=\,n/2\,$ or $J\,=\,(n-1)/2$ for the even or odd $n$ correspondingly and coefficients $C^{\,a_{1}\,\cdots\,a_{k}}_{b_{1}\,\cdots\,b_{p}}$ is a product of all
possible $\delta^{\,a_{i}\,b_{j}}$ delta functions which account all color indexes permutations in the expression.
Taking derivative of this expression in respect to the rapidity dependence of the Reggeon fields correlators, see \eq{Sec7},  we obtain
the evolution equation for the correlators of $\hat{O}$ operators as well:
\beqar\label{Sec13}
&\,& \frac{\D}{\D \eta}\,<\,\Le\,T^{a_1}\,\hat{O}_{1}(\tv_{\pm})\Ra\,\otimes\,\Le\,T^{a_2}\,\hat{O}_{2}(\tv_{\pm})\Ra\,\otimes\,\cdots\,\otimes\,
\Le\,T^{a_n}\,\hat{O}_{n}(\tv_{\pm})\Ra\,>\,\,= \,\nonumber \\
&=&C(R)^{n}\Le 2 g \Ra^{n}\int dx_{1}^{\pm}\cdots\int dx_{\,n}^{\pm}\Big(
\sum_{j=0}^{J} C_{b_1\cdots b_{n - 2j}}^{ a_1 \cdots a_{n}}\Le \frac{\imath}{2} \D^{-2}_{\bot} \Ra^{j} \Le -\imath \Ra^{n-2j}
\frac{\D}{\D \eta}
<\B_{\pm}^{\,b_1}\cdots \B_{\pm}^{\,b_{n-2j}}>\Big)
\eeqar
which determines the evolution of the Wilson line like operators of interests in terms of the effective vertices of Reggeon fields interactions, see \eq{Add1}, \eq{Sec5}-\eq{Sec7}.
We conclude also, that  due the structure of the Reggeon fields in the r.h.s of the expressions, see details in \cite{Our4}, the LO contribution to the correlators of the
$\hat{O}$ operators
is pure transverse in the quasi-multi-Regge kinematics for which the Lipatov's effective action is defined.

 Considering an equation for the correlator of two Wilson line operators and writing explicitly \eq{Sec121} for $n=2$ we obtain:
\beq\label{Sec14}
<tr\Le\,T^{a}\,\hat{O}(\tv_{+})\Ra\,\otimes\,tr\Le\,T^{b}\,\hat{O}(\tv_{-})\Ra>=4\,C(R)^{2}\,g^{2}\int dx^{+} \int dy^{-}\,\Le
\frac{\imath}{2}\, \hat{G}_{0\,-}^{\,a b\,+}-<\,\B_{+\,x}^{a}\,\B_{-\,y}^{b}>\Ra\,.
\eeq
Multiplying \eq{Sec14} on $T^{a}$ and $T^{b}$ matrices and using completeness identities for the matrices in the fundamental representation, we can rewrite \eq{Sec14} in the large $N_c$
limit as
\beq\label{Sec141}
<\,\hat{O}_{i j}(\tv_{+})\,\otimes\,\hat{O}_{k l}(\tv_{-})\,>\,=\,g^{2}\int dx^{+} \int dy^{-}\,\Le\,
\imath\,\delta_{i l}\,\delta_{j k}\, \hat{G}_{0\,-}^{\,+}\,-\,4\,<\Le \,\B_{+\,x}\Ra_{i j}\,\Le\,\B_{-\,y} \Ra_{k l}>\Ra\,.
\eeq
This expression is transverse to the leading order and in the shock wave approximation, where $\B_{\pm}\,=\,\delta(x^{\pm})\,\beta_{\pm}(x_{\bot})$
is assumed, the equation is simplified by the cancellation of the integration on $x^{+}$ and $y^{-}$  in it.
Therefore, we obtained that if we rewrite \eq{E16} in terms of the Wilson line correlators with the use of \eq{Sec141} expression
in the shock wave approximation then \eq{E16} can be represented as a JIMWLK-Balitsky like equation for the correlators of transverse Wilson
lines\footnote{We plan to rederive this equation in some separate publication, see also \cite{Hetch}}, see \cite{BK,Kovner,Hatta1}.
We also note, that in this case the corresponding Reggeon interaction vertices (kernels) in the equation must be taken in the
large $N_c$ limit in so called dipole approximation, see \cite{Fadin,LipIof}.

  Nevertheless, in general, $\,\B_{\pm}\,=\,\B_{\pm}(x^{\pm},x_{\bot})\,+\,D(x^{\pm},x^{\mp},x_{\bot})$ with $D$ as some corrections to the mean value of reggeized field $B$, see
\cite{Our4}, and therefore the \eq{Sec14} can not be precisely transverse in the full kinematical region of interest.
Additionally, due the complex structure of the all-order equation for the two-field correlator, see Appendix B,  the dependence on the $x^{-}$ and $y^{+}$ variables
in \eq{Sec14} can arises also through the family of the non-local vertexes $K$ in \eq{B3}, these corrections are beyond the shock wave and large $N_c$
approximations and in general will provide additional contributions for the correlators of Wilson lines as well. .

\section{Conclusion}

$\,\,\,\,\,\,$ In this paper we investigated two interrelated tasks: construction of a hierarchy of correlators of reggeized gluon fields in the formalism of Lipatov's effective action,
see \eq{Sec6}-\eq{Sec7} and Appendices A-D
 and connection
of these correlators to the correlators of operators of Wilson lines built from the longitudinal gluon fields, see \eq{Sec121}-\eq{Sec13}. These equations are main results of the paper
with the results obtained for the two Reggeon field correlator, see Section 3 and \eq{Sec14}-\eq{Sec141}. The equations for this correlator can be considered as analog of BK equation
derived in the formalism, it a new result in high energy QCD.

 Formulated as RFT, the Lipatov's approach to high energy scattering allows to derive the equations for the all types of correlators of the fields of reggeized gluons. Formally,
in the language of BFKL physics, each correlator represents some bound state of the Reggeons, i.e. \eq{B3} correlator, for example, represents a bound sate of two reggeized gluon fields
that corresponds to the propagator of reggeized gluons in regular BFKL calculus. Consequently, \eq{D1} correlator, is a bound state of four reggeized fields and represents BFKL Pomeron
like bound state which is after a suitable projection of color indexes will represent the BFKL Pomeron.
Other correlators, considered in the paper, are not widely used in the high energy scattering approaches but
they are important from the point of view of an accounting of the non-leading non-linear unitarity corrections to the scattering amplitudes, see
\cite{LipatovEff, BKP,GLR,BK,TripleV}.
The subsequent solution of the RFT equations of the hierarchy, see Appendixes A-D, with all vertices included, allows to determine the corrections to the leading poles
of the correlators in the momentum space in the way
different from the QCD perturbative scheme. The possible interconnection between the calculations of the kernels of interest in BFKL calculus,
see \cite{Fadin,Fadin1,LipIof},
and RFT calculus
it is a interesting problem which we we plan investigate in the future. We also note, that usual form of BFKL like evolution equations, see \cite{LipatovEff,BFKL,LipIof},
arises naturally in the RFT as a consequence of the dependence of the effective vertices of the theory on  the ultraviolet cut-off of the longitudinal momenta
in the expressions, see \eq{Sec7} and \eq{Sec71}.

  We note also, that the system of equations for the correlators \eq{Sec6}-\eq{Sec7} determines any correlator of interest in terms of
bare RFT propagator, see \eq{B4} or \eq{E6}, and vertices of the effective action only, which is regular property of Dyson-Schwinger hierarhy.
Therefore, for the different bare propagators chosen, the  formally different final answers will be obtained after the solution of the equations, which, nevertheless, must coincide in
the each particular perturbative order.
In turn, the vertices in the Lipatov's action are responsible for the account of the unitarity corrections to the amplitude
due the general non-conservation of the number of reggeized gluons in the scatering amplitude, see also \cite{BKP,GLR,BK,TripleV}.
In this case, quantum corrections to the amplitudes are given in the framework of RFT, whereas initially the expressions for the vertices are determined by perturbative QCD only.
Therefore, another important task for the future research  is a comparison
of the higher order perturbative corrections to the vertices arising from the QCD calculations, i.e. in BFKL formalism, with
the perturbative corrections to the same vertices determined by the Dyson-Schwinger hierarhy of the equtions derived in the
RFT calculus formalism.
Additionally it will be interesting to investigate the connection between correlators of two and four reggeized gluons.
It is known, that the Hamiltonian systems for the both bound states are integrable, see \cite{LipatovEff,LipIof,LipVega}. In the RFT formalism these states are not independent
and it is possible that some interconnection between the states on the language of integrable systems exists, see \cite{BP} for some discussion concerning this point.

 Basing on results of \cite{Our4}, we rewite the Lipatov's effective action in the form of the another effective action without reggeized gluons present.
The later action describes Wilson lines operators (Lipatov's operators) built from the longitudinal gluons interacting in two-dimensional transverse
plane with the help of corresponding two-dimensional propagator. An averaging of this interaction term over the gluon fields leads to the precise Lipatov's action
form, in which, nevertheless, the form of the corresponding Lipatov's operators can be different,
see \eq{Add4}-\eq{Add41} and \eq{WL1}-\eq{WL4} in the paper and remarks in \cite{Our4}.
The important advantage of the \eq{WL2} new action
is that it allows to connect the reggeized gluons correlators with the correlators of Wilson lines operators
built from gluons  beyond any simplifying approximations, see \eq{Sec121}-\eq{Sec13}. It is interesting to note also, that requesting Hermicity
of the Lipatov's operators, the form of the Lipatov's action will be different from the standard one to the non-leading orders, see \eq{Add41}.
That, in turn, affects on the form of non-leading corrections for the vertices in \eq{Sec1} that means the different
expressions in the r.h.s. of \eq{Sec121}-\eq{Sec13} which correspond to the different combinations of the Wilson lines in the correlators in the l.h.s. of the
equations. Namely, the correlators for the regular Wilson lines \eq{WL1} will be different from the correlators of the \eq{WL11} Lipatov's operators and this
difference is introduced by the different non-leading corrections to the effective vertices appearing in the correlators of  the Reggeon in the r.h.s. of the equations.

 Taking large $N_c$ and shock-wave approximations in \eq{Sec121} or \eq{Sec13},
the hierarchy similar to the  Balitsky-JIMWLK hierarchy of Wilson lines can be derived inside the framework of the formalism,
see \eq{Sec14}-\eq{Sec141} and \cite{Hetch}. We also obtained, that the knowledge of the
correlators of the reggeized gluon fields, i.e. vertices of \eq{Sec1} action, will allow to calculate any sub-leading correction in the corresponding hierarchy
taking large $N_c$ limit in the final expressions for the vertices, see \cite{Fadin}. Thereby, that connection between the different high energy QCD
approaches can be clarified that will help in further investigations of the subject.

 To conclude, we developed an approach for the calculation of the correlators of reggeized gluons based on the Dyson-Scwhinger derivation of the hierarchy of the correlators
in QFT. The equations obtained for the different correlators in the proposed RFT allow to investigate unitarity corrections to the amplitudes in BFKL physics  from
some new angle of view and hopefully will be useful for the calculation and verification of high order unitarity corrections in the high energy QCD.
The relation of these correlators to the  correlators of Wilson lines operators built from the longitudinal gluons is also clarified and results of this relation, we hope,
will help to understand general landscape of the high energy QCD approaches.

 One of the authors (S.B.)  greatly benefited from numerous discussions with M.Hentschinski. Both authors are grateful to M.Zubkov and A.Prygarin for useful discussions.


\newpage
\section*{Appendix A: Correlators of single Reggeon field}
\renewcommand{\theequation}{A.\arabic{equation}}
\setcounter{equation}{0}

 First of all, we determine the perturbation scheme in the calculations. Namely, in the equations of interests
any vertex $K^{a_{1}\cdots\,a_{n}}_{b_{1}\cdots b_{m}}$ is proportional to the $g^{n+m}$ power of the coupling constant. Nevertheless, we have in mind also, that
the leading high-energy asymptotic contributions to the correlators behave  as $\Le \,g^{2}\,N_{c}\, \ln\Le s/s_{0}\Ra\, \Ra^{n} $. Therefore, further
we will not write the power of the coupling constant in the fromt of the vertices  having in mind that in the equation's terms the
power of color factor $N_{c}$ is important as well as the corresponding power of $\ln(s/s_0)$, as leading we will call the contributions which behaves as
$\Le \,g^{2}\,N_{c}\, \ln\Le s/s_{0}\Ra\, \Ra^{n} $, see \cite{BFKL,BK,Kovner,Hatta1}.

 Now, we consider \eq{Sec1} and \eq{Sec3} together and obtain the following equations for the quantum $\B_{\pm}^{a}$ Reggeon field:
\beqar\label{A1}
&\,&\D^{2}_{\bot}\,<\B_{-}^{\,a}>\,=\,\Le K_{\,a_1}^{\,a}\Ra^{+}_{-}\,<\B_{-}^{\,a_1}>\,+\,\Le K^{\,a\,a_1}\Ra^{+ +}\,<\B_{+}^{\,a_1}>\,+\,
\Le K^{\,a\,}_{\,a_1\,a_{2}}\Ra^{+}_{- -}\,<\B_{-}^{\,a_1}\,\B_{-}^{a_2}>\,+\,\nonumber \\
&+&\,2\,\Le K^{\,a\,a_1}_{\,a_{2}}\Ra^{+ +}_{-}\,<\B_{+}^{\,a_1}\,\B_{-}^{a_2}>\,+\,
3\,\Le K^{\,a\,a_1\,a_{2}}\Ra^{+ + +}\,<\B_{+}^{\,a_1}\,\B_{+}^{a_2}>\,+\,\nonumber \\
&+&\,\Le K^{\,a}_{\,a_1\,a_{2}\,a_3}\Ra_{- - -}^{+}\,<\B_{-}^{\,a_1}\,\B_{-}^{a_2}\,\B_{-}^{a_3}>\,+\,
\,2\,\Le K^{\,a\,a_1}_{\,a_{2}\,a_3}\Ra_{- -}^{+ +}\,<\B_{+}^{\,a_1}\,\B_{-}^{a_2}\,\B_{-}^{a_3}>\,+\,\nonumber \\
&+&
3\,\Le K^{\,a\,a_1\,a_2}_{\,a_3}\Ra_{- }^{+ + +}\,<\B_{+}^{\,a_1}\,\B_{+}^{a_2}\,\B_{-}^{a_3}>\,+\,
\,4\,\Le K^{\,a\,a_1\,a_2\,a_3}\Ra^{+ + + +}\,<\B_{+}^{\,a_1}\,\B_{+}^{a_2}\,\B_{+}^{a_3}>\,+\,\cdots
\eeqar
and
\beqar\label{A2}
&\,&\D^{2}_{\bot}\,<\B_{+}^{\,a}>\, = \,\Le K_{\,a}^{a_1}\Ra^{+}_{-}\,<\B_{+}^{\,a_1}>\,+\,\Le K_{a\,a_1}\Ra_{- -}\,<\B_{-}^{\,a_1}>\,+\,
\Le K_{\,a\,}^{\,a_1\,a_{2}}\Ra^{+ +}_{-}\,<\B_{+}^{\,a_1}\,\B_{+}^{a_2}>\,+\,\nonumber \\
&+&\,2\,\Le K_{\,a\,a_2}^{\,a_{1}}\Ra^{+}_{- -}\,<\B_{+}^{\,a_1}\,\B_{-}^{a_2}>\,+\,
3\,\Le K_{\,a\,a_1\,a_{2}}\Ra_{- - -}\,<\B_{-}^{\,a_1}\,\B_{-}^{a_2}>\,+\,\nonumber \\
&+&\,\Le K_{\,a}^{\,a_1\,a_{2}\,a_3}\Ra_{-}^{+ + +}\,<\B_{+}^{\,a_1}\,\B_{+}^{a_2}\,\B_{+}^{a_3}>\,+\,
\,2\,\Le K_{\,a\,a_1}^{\,a_{2}\,a_3}\Ra_{- -}^{+ +}\,<\B_{-}^{\,a_1}\,\B_{+}^{a_2}\,\B_{+}^{a_3}>\,+\,\nonumber \\
&+&
3\,\Le K_{\,a\,a_1\,a_2}^{\,a_3}\Ra_{- - - }^{+}\,<\B_{-}^{\,a_1}\,\B_{-}^{a_2}\,\B_{+}^{a_3}>\,+\,
\,4\,\Le K_{\,a\,a_1\,a_2\,a_3}\Ra_{- - - -}\,<\B_{-}^{\,a_1}\,\B_{-}^{a_2}\,\B_{-}^{a_3}>\,+\,\cdots\,
\eeqar
\begin{figure}[!h]
\centering
a)
\includegraphics[scale=.5]{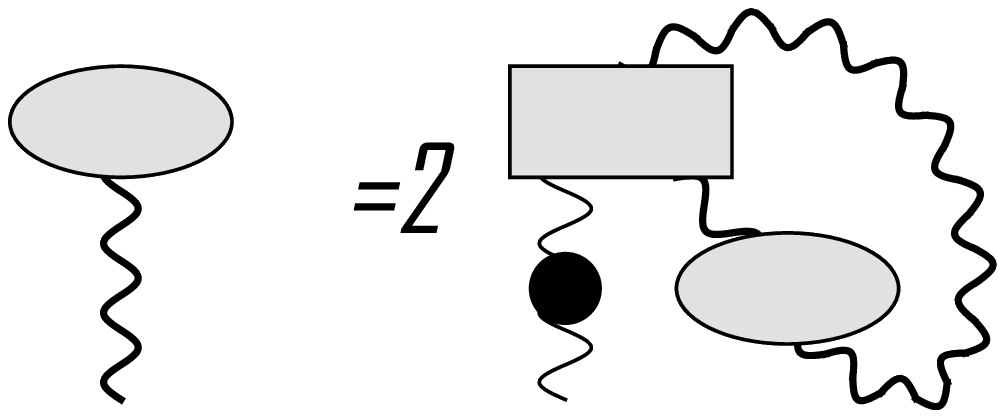}
\,\,\,\,\,\,\,\,\,\,
b)
\includegraphics[scale=.5]{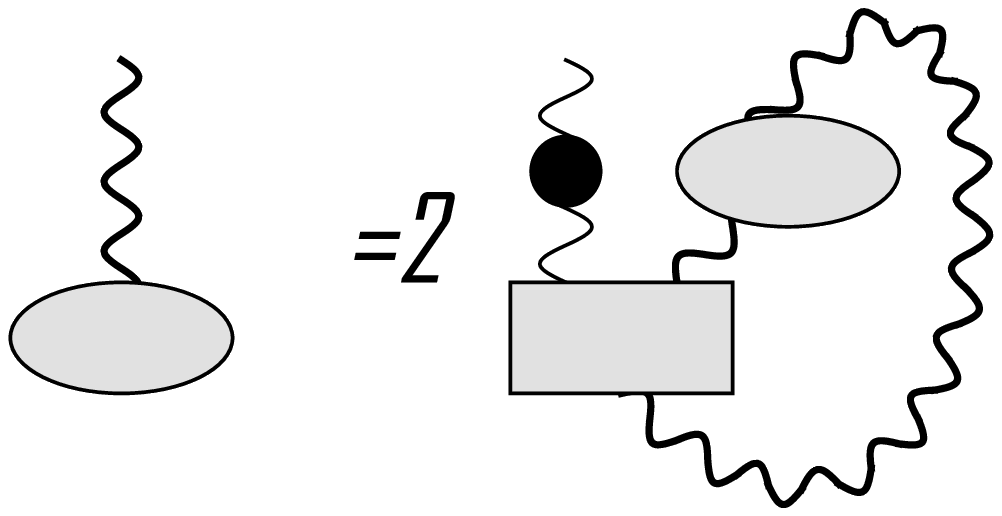}
\caption{
The diagrams a) and b) represent  \eq{A3} and  \eq{A4}. There are correlators $<B_{\pm}^{a_{1}} \cdots>$,  vertices $K^{a_{1}\cdots\,a_{n}}_{b_{1}\cdots b_{m}}$ and
  "bare" propagator of the Reggeon fields $G_{0\,-}^{\,b c\,+}$, introduced in \eq{B4}, denoted by ellipse, rectangle and point figures correspondingly in this and following diagrams.}
\label{fig:digraph}
\end{figure}

Now, using \eq{B4}-\eq{B5} expressions, we will get for the leading contributions to the quantum Reggeon fields:
\beqar\label{A3}
<\B_{-}^{\,a}>\,& = &\,2\,G_{0\,-}^{\,a\,a_{1}\,+}\,\Le K^{\,a_1\,a_2}_{\,a_{3}}\Ra^{+ +}_{-}\,< \B_{+}^{a_2} \B_{-}^{a_3}>\,=\,\nonumber \\
&=&
\,-\,2\,\imath\,G_{0\,-}^{\,a\,a_{1}\,+}\,\Le K^{\,a_1\,a_2}_{\,a_{3}}\Ra^{+ +}_{-}\,G_{0\,-}^{\,a_2\,a_{3}\,+}\,+
2\,G_{0\,-}^{\,a\,a_{1}\,+}\,\Le K^{\,a_1\,a_2}_{\,a_{3}}\Ra^{+ +}_{-}\,< \B_{+}^{a_2} \B_{-}^{a_3}>_1\,=\,\nonumber \\
&=&
\,-\,2\,\imath\,G_{0\,-}^{\,a\,a_{1}\,+}\,\Le K^{\,a_1\,a_2}_{\,a_{2}}\Ra^{+ +}_{-}\,G_{0}^{- +}\,+\,
2\,G_{0\,-}^{\,a\,a_{1}\,+}\,\Le K^{\,a_1\,a_2}_{\,a_{3}}\Ra^{+ +}_{-}\,< \B_{+}^{a_2} \B_{-}^{a_3}>_1\,
\eeqar
and
\beqar\label{A4}
<\B_{+}^{\,a}>\,& = & 2\,G_{0\,-}^{\,a\,a_{1}\,+}\,\Le K^{\,a_2}_{\,a_1\,a_{3}}\Ra^{+}_{- -}\,< \B_{+}^{a_2} \B_{-}^{a_3}>\,=\,\nonumber \\
&=&
\,-\,2\,\imath\,G_{0\,-}^{\,a\,a_{1}\,+}\,\Le K^{\,a_2}_{\,a_1\,a_{3}}\Ra^{+}_{- -}\,G_{0\,-}^{\,a_2\,a_{3}\,+}\,+
2\,G_{0\,-}^{\,a\,a_{1}\,+}\,\Le K^{\,a_2}_{\,a_1\,a_{3}}\Ra^{+}_{- -}\,< \B_{+}^{a_2} \B_{-}^{a_3}>_1\,=\,\nonumber \\
&=&\,-\,2\,\imath\,G_{0\,-}^{\,a\,a_{1}\,+}\,\Le K^{\,a_2}_{\,a_1\,a_{2}}\Ra^{+}_{- -}\,G_{0}^{- +}\,+
\,2\,G_{0\,-}^{\,a\,a_{1}\,+}\,\Le K^{\,a_2}_{\,a_1\,a_{3}}\Ra^{+}_{- -}\,< \B_{+}^{a_2} \B_{-}^{a_3}>_1\,.
\eeqar
These expressions contain tadpole contributions built from the Reggeon propagator, which is not enhanced due the presence of the rapidity ordering of the
Reggeon fields in the propagator, see \cite{LipatovEff, Our2}. Additionally, to leading order, these effective vertices are proportional
to $f_{a b c}$ anti-symmetrical structure constant, therefore further we take:
\beq\label{A5}
<\B_{-}^{\,a}>\,\approx\,0\,
\eeq
and
\beq\label{A6}
<\B_{+}^{\,a}>\,\approx\,0\,,
\eeq
that corresponds  to the signature conservation law as well. We also note, that the coefficient $2$ in the front of \eq{A3}-\eq{A4} arises due the fact we do not symmetrize
the expressions in respect to $a_1$ and $a_2$ indexes. We will keep these form of the shorthand notations further in all the places where it will not lead to any confusion and will write the
precise symmetric expressions in respect to the indexes where it will be important.

\newpage
\section*{Appendix B: Correlators of two Reggeon fields}
\renewcommand{\theequation}{B.\arabic{equation}}
\setcounter{equation}{0}

 The $<\B_{+}^{\,a}\,\B_{-}^{\,a_1}>$ correlator of the Reggeon fields is corresponding to the propagator of the reggeized gluons, whereas
$<\B_{+}^{\,a}\,\B_{+}^{\,a_1}>$ and $<\B_{-}^{\,a}\,\B_{-}^{\,a_1}>$ correlators are suppressed perturbatively in comparison to the first one and represent
some kind of the "mass" terms of the Reggeon fields. Therefore, in the expression for the last two propagators only the first terms will be presented.
We have for the correlators of these fields:
\beq\label{B1}
\D^{2}_{\bot}\,<\B_{-}^{\,a}\,\B_{-}^{\,a_1}>\,=\,\Le K^{\,a}_{\,a_2} \Ra^{+}_{-}\,<\B_{-}^{\,a_2}\,\B_{-}^{\,a_1}>\,+\,2\,
\Le K^{\,a\,a_2} \Ra^{+ +}\,<\B_{+}^{\,a_2}\,\B_{-}^{\,a_1}>\,+\,\cdot\,
\eeq
and correspondingly
\beq\label{B2}
\D^{2}_{\bot}\,<\B_{+}^{\,a}\,\B_{+}^{\,a_1}>\,=\,\Le K_{a}^{\,a_2} \Ra^{+}_{-}\,<\B_{+}^{a_2}\,\B_{+}^{a_1}>\,+\,2\,
\Le K_{a\,a_2} \Ra_{- -}\,<\B_{-}^{a_2}\,\B_{+}^{a_1}>\,+\,\cdots\,.
\eeq
For the correlator of $\pm$ Reggeon fields we obtain:
\beqar\label{B3}
&\,&\D^{2}_{\bot}\,<\B_{+}^{\,a}\,\B_{-}^{\,a_1}>\,=\,-\,\imath\,\delta^{\,a a_1}\,+\,\Le K_{\,a}^{\,a_2} \Ra^{+}_{-}\,<\B_{+}^{\,a_2}\,\B_{-}^{\,a_1}>\,+\,
2\,\Le K_{a\,a_2} \Ra_{- -}\,<\B_{-}^{\,a_2}\,\B_{-}^{\,a_1}>\,+\,\nonumber \\
&+&\,\Le K^{\,a_2\,a_3}_{a}\Ra^{+ +}_{-}\,<\B_{+}^{\,a_3}\,\B_{+}^{a_2}\,\B_{-}^{a_1}>\,+\,2\,
\Le K^{\,a_2}_{a\,a_3}\Ra^{+\,}_{- -}\,<\B_{+}^{\,a_2}\,\B_{-}^{a_3}\,\B_{-}^{a_1}>\,+\,\nonumber \\
&+&\,3\,\Le K_{a\,a_2\,a_3}\Ra_{- - -}\,<\B_{-}^{\,a_3}\,\B_{-}^{a_2}\,\B_{-}^{a_1}>\,+\,
\Le K_{a}^{\,a_2\,a_3\,a_4}\Ra_{- }^{+ + +}\,<\B_{+}^{\,a_2}\,\B_{+}^{a_3}\,\B_{+}^{a_4}\,\B_{-}^{\,a_1}>\,+\,\nonumber \\
&+&\,2\,\Le K_{a\,a_2}^{\,a_3\,a_4}\Ra_{- - }^{+ +}\,<\B_{+}^{\,a_3}\,\B_{+}^{a_4}\,\B_{-}^{a_2}\,\B_{-}^{\,a_1}>\,+\,
3\,\Le K_{a\,a_2\,a_3}^{\,a_4}\Ra_{- - - }^{+}\,<\B_{+}^{\,a_4}\,\B_{-}^{a_3}\,\B_{-}^{a_2}\,\B_{-}^{\,a_1}>\,+\,\nonumber \\
&+&\,4\,\Le K_{a\,a_2\,a_3\,a_4}\Ra_{- - - - }\,<\B_{-}^{\,a_4}\,\B_{-}^{a_3}\,\B_{-}^{a_2}\,\B_{-}^{\,a_1}>\,+\,\cdots\,.
\eeqar
\begin{figure}[!h]
\centering
\includegraphics[scale=.9]{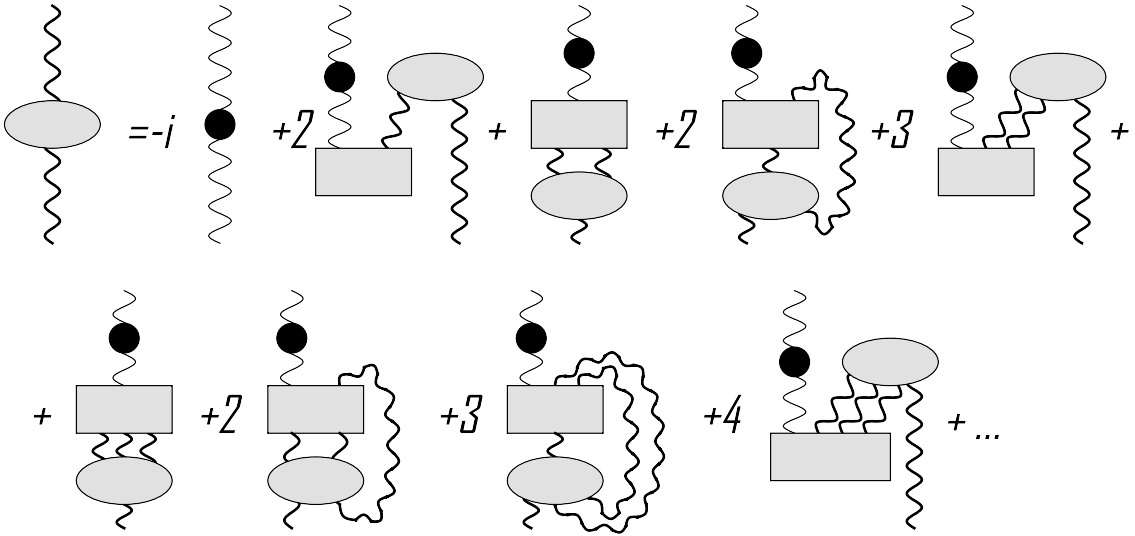}
\caption{The diagram represents  \eq{B3}. It denotes equation for $<\B_{+}^{\,a}\,\B_{-}^{\,a_1}>$ correlator. }
\label{fig:digraph1}
\end{figure}

Now we can introduce the "bare" propagator of the Reggeon fields as
\beq\label{B4}
\Le\,\delta^{a\,b} \D^{2}_{\bot}\,-\,\Le K^{\,a}_{\,b} \Ra^{+}_{-}\,\Ra\, G_{0\,-}^{\,b c\,+}\,=\,\delta^{a\,c}
\eeq
and obtain the following perturbative expression for the correlator:
\beq\label{B5}
<\B_{+}^{\,a} \B_{-}^{\,b}>=<\B_{+}^{\,a} \B_{-}^{\,b}>_{0}+<\B_{+}^{\,a} \B_{-}^{\,b}>_{1}=-\imath G_{0\,-}^{\,a b\,+}+
<\B_{+}^{\,a} \B_{-}^{\,b}>_{1}=-\imath\,\delta^{a b} G_{0}^{+ -}+ <\B_{+}^{\,a}\,\B_{-}^{\,b}>_{1}\,
\eeq
with
\beq
<\B_{+}^{\,a}\,\B_{-}^{\,b}>_{1}\,\propto\,\delta^{\,a b}
\eeq
as well.
Therefore, we have for the leading contributions to the r.h.s. of \eq{B1} and \eq{B2}:
\beqar\label{B6}
<\B_{-}^{\,a}\,\B_{-}^{\,b}>\,&=&\,2\,G_{0\,-}^{\,a a_1\,+}\,\Le K^{\,a_1\,a_2} \Ra^{+ +}\,<\B_{+}^{a_2} \B_{-}^{b}>\,=\,\nonumber\\
\,&-&\,2\,\imath\,G_{0\,-}^{\,a a_1\,+}\,\Le K^{\,a_1\,a_2} \Ra^{+ +}\,G_{0\,-}^{\,a_2 b\,+}\,+\,
2\,G_{0\,-}^{\,a a_1\,+}\,\Le K^{\,a_1\,a_2} \Ra^{+ +}\,<\B_{+}^{a_2} \B_{-}^{b}>_1
\eeqar
and
\beqar\label{B7}
<\B_{+}^{\,a}\,\B_{+}^{\,b}>\,&=&\,2\,G_{0\,-}^{\,a a_1\,+}\,\Le K_{\,a_1\,a_2} \Ra_{- -}\,<\B_{+}^{b} \B_{-}^{a_2}>=\,\nonumber\\
&-&\,2\,\imath\,G_{0\,-}^{\,a a_1\,+}\,\Le K_{\,a_1\,a_2} \Ra_{- -}\,G_{0\,-}^{\,b a_2\,+}\,+\,
2\,G_{0\,-}^{\,a a_1\,+}\,\Le K_{\,a_1\,a_2} \Ra_{- -}\,<\B_{+}^{b} \B_{-}^{a_2}>_1\,.
\eeqar
Now, using \eq{B4}, \eq{B3} reads as:
\beqar\label{B8}
&\,&<\B_{+}^{\,a}\,\B_{-}^{\,a_1}>_1\,=\,
2\,G_{0\,-}^{\,a b\,+}\,\Le K_{b\,b_1} \Ra_{- -}\,<\B_{-}^{\,b_1}\,\B_{-}^{\,a_1}>\,+\,
G_{0\,-}^{\,a b\,+}\,\Le K^{\,b_1\,b_2}_{b}\Ra^{+ +}_{-}\,<\B_{+}^{\,b_1}\,\B_{+}^{b_2}\,\B_{-}^{a_1}>\,+\,\nonumber \\
&+&\,
2\,G_{0\,-}^{\,a b\,+}\,\Le K^{\,b_1}_{b\,b_2}\Ra^{+\,}_{- -}\,<\B_{+}^{\,b_1}\,\B_{-}^{b_2}\,\B_{-}^{a_1}>\,+\,
3\,G_{0\,-}^{\,a b\,+}\,\Le K_{b\,b_1\,b_2}\Ra_{- - -}\,<\B_{-}^{\,b_1}\,\B_{-}^{b_2}\,\B_{-}^{a_1}>\,+\,\nonumber \\
&+&\,
2\,G_{0\,-}^{\,a b\,+}\,\Le K_{b\,b_1}^{\,b_2\,b_3}\Ra_{- - }^{+ +}\,<\B_{+}^{\,b_2}\,\B_{+}^{b_3}\,\B_{-}^{b_1}\,\B_{-}^{\,a_1}>\,.
\eeqar
Inserting LO values of \eq{B6}, \eq{C5}-\eq{C6} and \eq{D3} expressions back into the \eq{B3}, we obtain the following NLO correction to the LO answer for the two Reggeon fields correlator:
\beqar\label{B9}
<\B_{+}^{\,a}\,\B_{-}^{\,a_1}>_1\,& = &\,-\,4\,\imath\,G_{0\,-}^{\,a b\,+}\,\Le K_{b\,b_1} \Ra_{- -}\,G_{0\,-}^{\,b_1 b_2\,+}\,\Le K^{\,b_2\,b_3} \Ra^{+ +}\,G_{0\,-}^{\,b_3 a_1\,+}\,
-\,\nonumber\\
&-&\,4\,G_{0\,-}^{\,a b\,+}\,\Le K^{\,b_1\,b_2}_{b}\Ra^{+ +}_{-}\,
G_{0\,-}^{\,b_1 a_1\,+}\,G_{0\,-}^{\,b_2 b_3\,+}\,\Le K^{\,b_5}_{\,b_3\,b_4}\Ra^{+}_{- -}\,G_{0\,-}^{\,b_5 b_4\,+}\,
-\nonumber \\
&-&\,6\,G_{0\,-}^{\,a b\,+}\,\Le K^{\,b_1\,b_2}_{b}\Ra^{+ +}_{-}\,G_{0\,-}^{\,b_4 b_{2}\,+}\,
G_{0\,-}^{\,b_1 b_3\,+}\,\Le K^{\, b_5}_{b_3\, b_4} \Ra^{ +}_{- -}\,G_{0\,-}^{\,b_5 a_{1}\,+}\,-\,\nonumber \\
&-&
4\,G_{0\,-}^{\,a b\,+}\,\Le K^{\,b_1}_{b\,b_2}\Ra^{+\,}_{- -}\,G_{0\,-}^{\,b_1 b_2\,+}\,G_{0\,-}^{\,a_1 b_3\,+}\,
\Le K^{\,b_3\,b_4 }_{b_5}\Ra^{+ +}_{-}\,G_{0\,-}^{\,b_4 b_5\,+}-\,\nonumber \\
&-&
4\,G_{0\,-}^{\,a b\,+}\,\Le K^{\,b_1}_{b\,b_2}\Ra^{+\,}_{- -}\,G_{0\,-}^{\,b_1 a_1\,+}\,G_{0\,-}^{\,b_2 b_3\,+}\,
\Le K^{\,b_3\,b_4 }_{b_5}\Ra^{+ +}_{-}\,G_{0\,-}^{\,b_4 b_5\,+}\,-\,\nonumber \\
&-&
18\,G_{0\,-}^{\,a b\,+}\,\Le K_{b\,b_1\,b_2}\Ra_{- - -}\,G_{0\,-}^{\,b_3 b_1\,+}\,G_{0\,-}^{\,b_5 b_2\,+}\,\Le K^{\,b_3\,b_4\,b_5}\Ra^{+ + +}\,
G_{0\,-}^{\,b_4 a_1\,+}\,\,-\,\nonumber \\
&-&
4\,G_{0\,-}^{\,a b\,+}\,\Le K_{b\,b_1}^{\,b_2\,b_3}\Ra_{- - }^{+ +}\,G_{0\,-}^{\,b_2 b_1\,+}\,G_{0\,-}^{\,b_3 a_1\,+}\,.
\eeqar

\begin{figure}[!hb]
\centering
\includegraphics[scale=1]{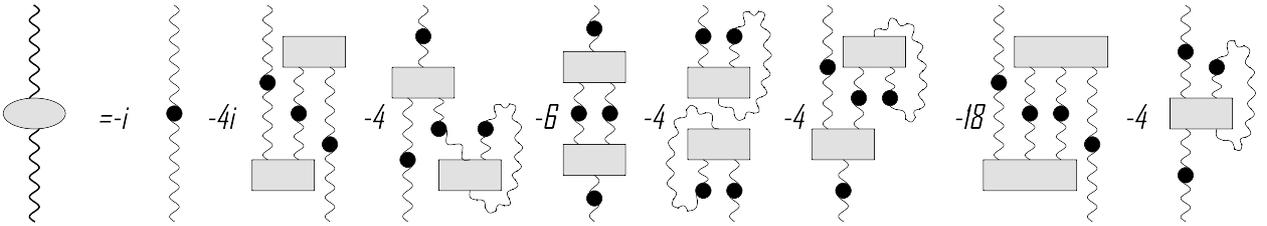}
\caption{The diagram represents \eq{B9}. It denotes perturbative solution for $<\B_{+}^{\,a}\,\B_{-}^{\,a_1}>$ correlator.}
\label{fig:digraph2}
\end{figure}
Comparing the different contributions in the  r.h.s. of the expression we see that the main contribution to the correlator comes from the last term in r.h.s. of \eq{B9}.

\newpage
\section*{Appendix C: Correlators of three Reggeon fields}
\renewcommand{\theequation}{C.\arabic{equation}}
\setcounter{equation}{0}

There are the following correlators of the three Reggeon fields which we have to calculate, the first one is the following.
\beqar\label{C1}
&\,&\D^{2}_{\bot}\,<\,\B_{+}^{\,a} \B_{+}^{\,a_1} \B_{-}^{\,a_2}\,>\, = \,-\,\imath\,\delta^{\,a a_2}\,< \B_{+}^{\,a_1}>\,+\,
\Le K^{\,a_3}_{a} \Ra^{+}_{-} <\,\B_{+}^{\,a_3} \B_{+}^{\,a_1} \B_{-}^{\,a_2}\,>\,+\,\nonumber \\
&+&2\,\Le K_{a a_3} \Ra_{- -} <\,\B_{-}^{\,a_3} \B_{+}^{\,a_1} \B_{-}^{\,a_2}\,>\,+
\Le K^{\,a_3 a_4}_{a} \Ra^{+ +}_{-} <\,\B_{+}^{\,a_3} \B_{+}^{a_4} \B_{+}^{\,a_1} \B_{-}^{\,a_2}\,> + \nonumber \\
&+&2 \Le K^{\,a_4}_{a a_3} \Ra^{+}_{- -} <\,\B_{+}^{\,a_4} \B_{-}^{a_3} \B_{+}^{\,a_1} \B_{-}^{\,a_2}\,> +
3 \Le K_{a a_3 a_4} \Ra_{- - -} <\,\B_{-}^{\,a_3} \B_{-}^{a_4} \B_{+}^{\,a_1} \B_{-}^{\,a_2}\,>\,.
\eeqar
The second one can be obtained from the first one by replace $+$ on $-$ in the expression:
\beqar\label{C2}
&\,&\D^{2}_{\bot}\,<\,\B_{-}^{\,a} \B_{-}^{\,a_1} \B_{+}^{\,a_2}\,>\, = \,-\,2 \imath\,\delta^{\,a a_2}\,< \B_{-}^{\,a_1}>\,+\,
\Le K^{\,a}_{a_3} \Ra^{+}_{-} <\,\B_{-}^{\,a_3} \B_{+}^{\,a_1} \B_{+}^{\,a_2}\,>\,+\,\nonumber \\
&+& 2 \,\Le K^{a a_3} \Ra^{+ +} <\,\B_{+}^{\,a_3} \B_{-}^{\,a_1} \B_{+}^{\,a_2}\,>\,+
\Le K_{\,a_3 a_4}^{a} \Ra^{+}_{- -} <\,\B_{-}^{\,a_3} \B_{-}^{a_4} \B_{-}^{\,a_1} \B_{+}^{\,a_2}\,> + \nonumber \\
&+& 2 \Le K_{\,a_4}^{\,a a_3} \Ra^{+ +}_{-} <\,\B_{-}^{\,a_4} \B_{+}^{a_3} \B_{-}^{\,a_1} \B_{+}^{\,a_2}\,> +
3 \Le K^{\,a a_3 a_4} \Ra^{+ + +} <\,\B_{+}^{\,a_3} \B_{+}^{a_4} \B_{-}^{\,a_1} \B_{+}^{\,a_2}\,>\,
\eeqar
and the third one:
\beqar\label{C3}
&\,&\D^{2}_{\bot}\,<\,\B_{-}^{\,a} \B_{-}^{\,a_1} \B_{-}^{\,a_2}\,>\, =\, \Le K_{a_3}^{\,a} \Ra^{+}_{-} <\,\B_{-}^{\,a_3} \B_{-}^{\,a_1} \B_{-}^{\,a_2}\,>\,+\,
2 \Le K^{\,a a_3} \Ra^{+}_{-} <\,\B_{+}^{\,a_3} \B_{-}^{\,a_1} \B_{-}^{\,a_2}\,>\,+\,\nonumber \\
&+& \Le K^{\,a}_{a_3 a_4} \Ra^{+}_{- -} <\,\B_{-}^{\,a_3} \B_{-}^{a_4} \B_{-}^{\,a_1} \B_{-}^{\,a_2}\,>\,+\,
2 \Le K^{\,a a_3}_{a_4} \Ra^{+ +}_{-} <\,\B_{+}^{\,a_3} \B_{-}^{a_4} \B_{-}^{\,a_1} \B_{-}^{\,a_2}\,>\,+\,\nonumber \\
&+& 3 \Le K^{\,a a_3 a_4} \Ra^{+ + +} <\,\B_{+}^{\,a_3} \B_{+}^{a_4} \B_{-}^{\,a_1} \B_{-}^{\,a_2}\,>\,.
\eeqar
This system of equations can be solved perturbatively, using results of the Appendix A and Appendix D, here we will use the symmetric expression for \eq{D3}:
\beqar\label{C31}
<\,\B_{+}^{\,a} \B_{+}^{\,a_1} \B_{-}^{\,a_2} \B_{-}^{\,a_3} \,>\,& = & \,-\,G_{0\,-}^{\,a a_{3}\,+}\,G_{0\,-}^{\,a_1 a_{2}\,+}\,-\,
G_{0\,-}^{\,a a_{2}\,+}\,G_{0\,-}^{\,a_1 a_{3}\,+}\,-\,\nonumber \\
&-&\,\imath\,G_{0\,-}^{\,a a_{3}\,+}\,<\,\B_{+}^{\,a_1} \B_{-}^{a_2} \,>_1\,-\,
\imath\,G_{0\,-}^{\,a_1\, a_{2}\,+}\,<\,\B_{+}^{\,a} \B_{-}^{a_3} \,>_1\,.
\eeqar
We obtain for \eq{C3}:
\beqar\label{C4}
&\,& <\,\B_{-}^{\,a} \B_{-}^{\,a_1} \B_{-}^{\,a_2}\,>\,  = \, 2\, G_{0\,-}^{\,a b\,+}\,\Le K^{\,b b_1} \Ra^{+}_{-} <\,\B_{+}^{\,b_1} \B_{-}^{\,a_1} \B_{-}^{\,a_2}\,>
\,-\, \nonumber\\
&-& 6 \,G_{0\,-}^{\,a \B_2\,+}\, \Le K^{\,b_2 b_3 b_4} \Ra^{+ + +}
\Big( G_{0\,-}^{\,b_3 a_{2}\,+}\,G_{0\,-}^{\,b_4 a_{1}\,+} +
\imath\, G_{0\,-}^{\,b_3 a_{2}\,+}\,<\,\B_{+}^{\,b_4} \B_{-}^{a_1} \,>_1 \Big)\,,
\eeqar
that to leading order gives
\beqar\label{C41}
<\,\B_{-}^{\,a} \B_{-}^{\,a_1} \B_{-}^{\,a_2}\,>\,&  = & \,-\,6 \imath\, G_{0\,-}^{\,b_1 a_{2}\,+}\,\Le K^{\,b b_1 b_2} \Ra^{+ + +}\, G_{0\,-}^{\,b_1 a_{2}\,+}\,
<\,\B_{+}^{\,b_2} \B_{-}^{\,a_1} \,>\,=\,\nonumber \\
&=&
-6 \,G_{0\,-}^{\,a b\,+}\, \Le K^{\,b b_1 b_2} \Ra^{+ + +}
\Big( G_{0\,-}^{\,b_1 a_{2}\,+}\,G_{0\,-}^{\,b_2 a_{1}\,+} +
\imath G_{0\,-}^{\,b_1 a_{2}\,+}\,<\,\B_{+}^{\,b_2} \B_{-}^{\,a_1} \,>_1 \Big).
\eeqar
Correspondingly, for \eq{C1} we have to the leading order:
\beqar\label{C5}
&\,&<\,\B_{+}^{\,a} \B_{+}^{\,a_1} \B_{-}^{\,a_2}\,>\, = \,-\,\imath\,G_{0\,-}^{\,a a_{2}\,+}\,\,< \B_{+}^{\,a_1}>\,+\,
2\,G_{0\,-}^{\,a b\,+}\, \Le K^{\,b_1}_{b b_2} \Ra^{+}_{- -} <\,\B_{+}^{\,b_1}  \B_{+}^{\,a_1} \B_{-}^{\,a_2} \B_{-}^{b_2}\,>\,= \nonumber \\
&=&\,
-\,2\,\imath\,G_{0\,-}^{\,a a_2\,+}\,G_{0\,-}^{\,a_1\,b\,+}\,\Le K^{\,b_1}_{b\,\,b_{2}}\Ra^{+ }_{- -}\,<\B_{+}^{b_1} \B_{-}^{b_2}>
\,-\,2\imath\,G_{0\,-}^{\,a_1\,b\,+}\,\Le K^{\,b_1}_{b\,\,b_{2}}\Ra^{+ }_{- -}\,G_{0\,-}^{\,b_1 b_2\,+}\,<\B_{+}^{a} \B_{-}^{a_2}>\,-\nonumber \\
&-&\imath\,G_{0\,-}^{\,a b\,+}\,\Le K^{\, b_1}_{b\, b_2} \Ra^{+ }_{- -}
G_{0\,-}^{\,b_1 a_{2}\,+}\,<\B_{+}^{\,a_1} \B_{-}^{\,b_2} >\,
-
\imath\,G_{0\,-}^{\,a b\,+}\,\Le K^{\, b_1}_{b\, b_2} \Ra^{+ }_{- -}
G_{0\,-}^{\,b_2 a_{2}\,+}\,<\B_{+}^{\,b_1} \B_{-}^{\,a_1} >
= \nonumber \\
&=&\,
\,-4\,G_{0\,-}^{\,a a_2\,+}\,G_{0\,-}^{\,a_1\,b\,+}\,\Le K^{\,b_1}_{b\,\,b_{2}}\Ra^{+}_{- -}\,G_{0\,-}^{\,b_1\,b_{2}\,+}\,-\,
2\,\imath\,G_{0\,-}^{\,a a_2\,+}\,G_{0\,-}^{\,a_1\,b\,+}\,\Le K^{\,b_1}_{b\,\,b_{2}}\Ra^{+ }_{- -}\,<\B_{+}^{b_1} \B_{-}^{b_2}>_{1}\,-\nonumber \\
&-&
2\,G_{0\,-}^{\,a b\,+}\,\Le K^{\, b_2}_{b\, b_1} \Ra^{ +}_{- -}\,G_{0\,-}^{\,a_1 b_{1}\,+}\,G_{0\,-}^{\,b_2 a_{2}\,+}\,-\,
\,2\imath\,G_{0\,-}^{\,a_1\,b\,+}\,\Le K^{\,b_1}_{b\,\,b_{2}}\Ra^{+ }_{- -}\,G_{0\,-}^{\,b_1 b_2\,+}\,<\B_{+}^{a} \B_{-}^{a_2}>_{1}-\nonumber \\
&-&
\imath\,G_{0\,-}^{\,a b\,+}\,\Le K^{\, b_1}_{b\, b_2} \Ra^{+ }_{- -}
G_{0\,-}^{\,b_1 a_{2}\,+}\,<\B_{+}^{\,a_1} \B_{-}^{\,b_2} >_1 -
\imath\,G_{0\,-}^{\,a b\,+}\,\Le K^{\, b_1}_{b\, b_2} \Ra^{+ }_{- -}
G_{0\,-}^{\,b_2 a_{1}\,+}\,<\B_{+}^{\,b_1} \B_{-}^{\,a_2} >_1\,,
\eeqar
\begin{figure}[!t]
\centering
\includegraphics[scale=0.7]{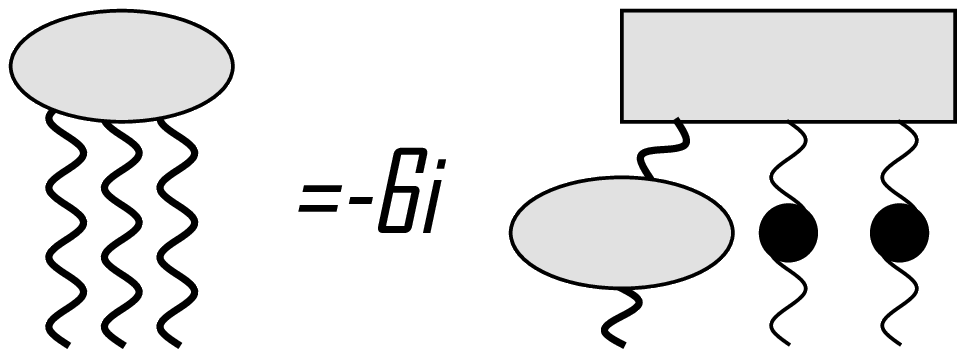}
\caption{The represents \eq{C4}. It denotes LO expression of  $<\,\B_{-}^{\,a} \B_{-}^{\,a_1} \B_{-}^{\,a_2}\,>$ correlator in terms of Reggeon propagator.}
\label{fig:digraph3}
\end{figure}
\begin{figure}[!t]
\centering
\includegraphics[scale=1]{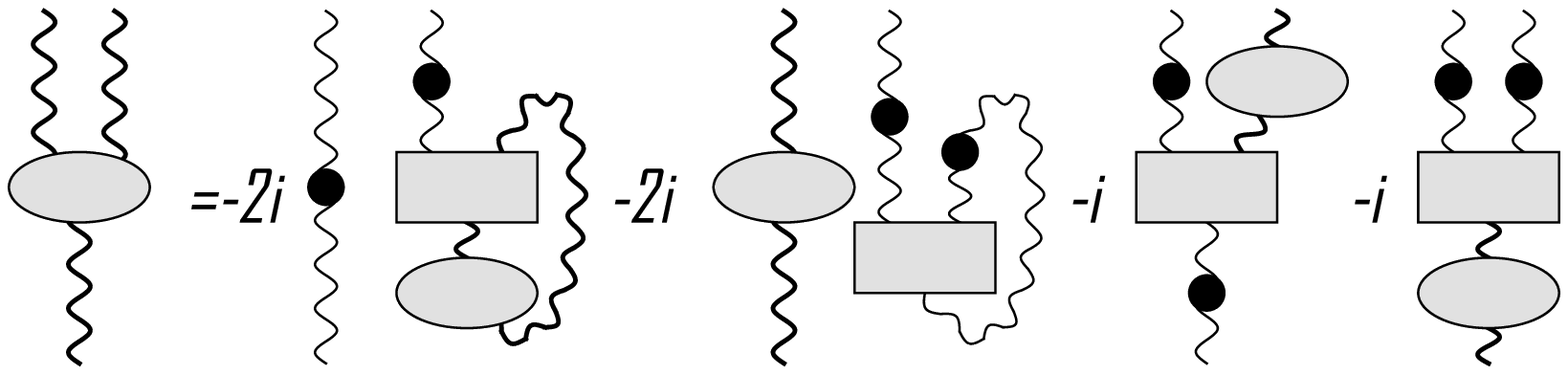}
\caption{The represents  \eq{C5}. It denotes LO expression of $<\,\B_{+}^{\,a} \B_{+}^{\,a_1} \B_{-}^{\,a_2}\,>$ correlator in terms of Reggeon propagator.}
\label{fig:digraph4}
\end{figure}
\begin{figure}[!t]
\centering
\includegraphics[scale=1]{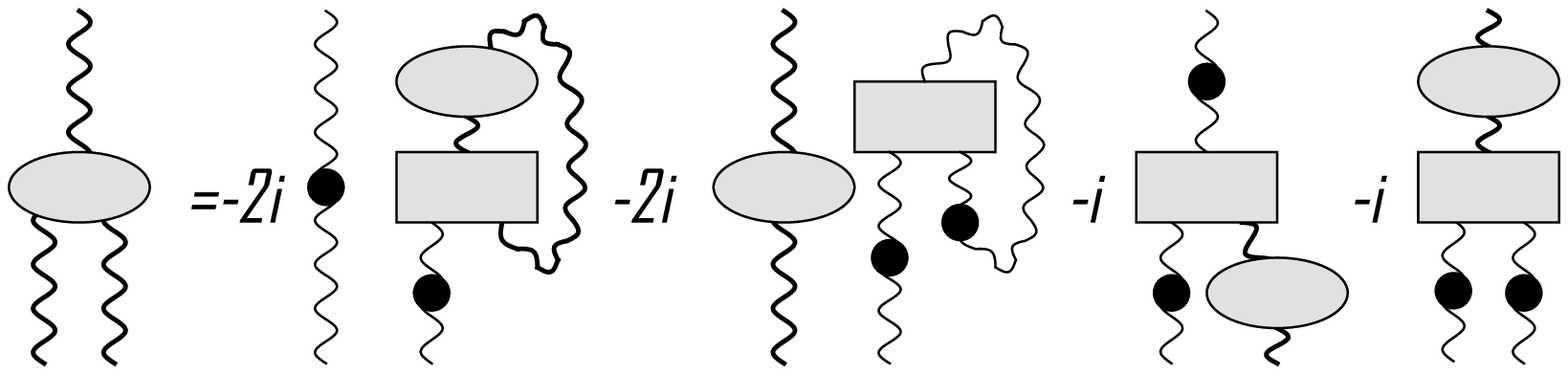}
\caption{The diagram represents \eq{C6}. It denotes LO expression of  $<\,\B_{+}^{\,a} \B_{-}^{\,a_1} \B_{-}^{\,a_2}\,>$ correlator in terms of Reggeon propagator.}
\label{fig:digraph5}
\end{figure}
we note, that we wrote \eq{C5} in the mostly symmetrical way.
Correspondingly, the answer for \eq{C2} can be obtained from \eq{C5} by replace of the $+$ and $-$ signs in \eq{C5}:
\beqar\label{C6}
&\,&<\,\B_{+}^{\,a} \B_{-}^{\,a_1} \B_{-}^{\,a_2}\,>\, = \,-\,2 \imath\,G_{0\,-}^{\,a a_1\,+}\,< \B_{-}^{\,a_2}>\,+\,G_{0\,-}^{\,a b\,+}\,
\Le K^{\,b_1 b_2}_{b} \Ra^{+ +}_{-} <\,\B_{+}^{\,b_1} \B_{+}^{\,b_2} \B_{-}^{\,a_1} \B_{-}^{\,a_2}\,>\,= \nonumber \\
&=&\,
\,-\,2\,\imath\,G_{0\,-}^{\,a a_1\,+}\,G_{0\,-}^{\,a_2\,b\,+}\,\Le K^{\,b\,b_1}_{\,b_{2}}\Ra^{+ +}_{-}\,<\B_{+}^{b_1} \B_{-}^{b_2}>\,-
2\,\imath\,G_{0\,-}^{\,a_2\,b\,+}\,\Le K^{\,b\,b_1}_{\,b_{2}}\Ra^{+ +}_{-}\,G_{0\,-}^{\,b_1 b_2\,+}\,<\B_{+}^{a} \B_{-}^{a_1}>\,-\,\nonumber \\
&-&\imath\,G_{0\,-}^{\,a b\,+}\,\Le K^{\,b_1 b_2}_{b} \Ra^{+ +}_{-}\,
G_{0\,-}^{\,b_1 a_{2}\,+}\,<\,\B_{+}^{\,b_2} \B_{-}^{\,a_1} \,> \,- \,\imath\,G_{0\,-}^{\,b_1 a_1\,+}\,\Le K^{\,b_1 b_2}_{b} \Ra^{+ +}_{-}\,
G_{0\,-}^{\,b_2 a_{2}\,+}\,<\,\B_{+}^{\,a} \B_{-}^{\,b} \,>
= \nonumber \\
&=&\,-4\,G_{0\,-}^{\,a a_1\,+}\,G_{0\,-}^{\,a_2\,b\,+}\,\Le K^{\,b\,b_1}_{\,b_{2}}\Ra^{+ +}_{-}\,G_{0\,-}^{\,b_1\,b_{2}\,+}\,-\,
2\,\imath\,G_{0\,-}^{\,a a_1\,+}\,G_{0\,-}^{\,a_2\,b\,+}\,\Le K^{\,b\,b_1}_{\,b_{2}}\Ra^{+ +}_{-}\,<\B_{+}^{b_1} \B_{-}^{b_2}>_{1}\,-\nonumber \\
&-&\,2\,G_{0\,-}^{\,a b\,+}\,\Le K^{\,b_1 b_2}_{b} \Ra^{+ +}_{-}\,G_{0\,-}^{\,b_1 a_{2}\,+}\,G_{0\,-}^{\,b_2 a_{1}\,+}\,-
2\,\imath\,G_{0\,-}^{\,a_2\,b\,+}\,\Le K^{\,b\,b_1}_{\,b_{2}}\Ra^{+ +}_{-}\,G_{0\,-}^{\,b_1 b_2\,+}\,<\B_{+}^{a} \B_{-}^{a_1}>_{1}\,-\,\nonumber \\
&-&2\,\imath\,G_{0\,-}^{\,a b\,+}\,\Le K^{\,b_1 b_2}_{b} \Ra^{+ +}_{-}\,
G_{0\,-}^{\,b_1 a_{2}\,+}\,<\,\B_{+}^{\,b_2} \B_{-}^{\,a_1} \,>_1 \,,
\eeqar
that can be  verified
by the direct calculation of \eq{C2}.

\newpage
\section*{Appendix D: Correlators of four Reggeon fields}
\renewcommand{\theequation}{D.\arabic{equation}}
\setcounter{equation}{0}

 We limit the chain of the equations by the 4-Reggeon correlators, therefore the equations for these correlators are simple.
We have for the symmetrical correlator of four Reggeon fields:
\beqar\label{D1}
\D^{2}_{\bot}\,<\,\B_{+}^{\,a} \B_{+}^{\,a_1} \B_{-}^{\,a_2} \B_{-}^{\,a_3} \,>\, &= &\,-\,2\,\imath\,\delta^{\,a a_3}\,<\,\B_{+}^{\,a_1} \B_{-}^{a_2} \,>\,+\,
\Le K^{\,a_4}_{a} \Ra^{+}_{-} <\,\B_{+}^{\,a_4} \B_{+}^{\,a_1} \B_{-}^{\,a_2} \B_{-}^{\,a_3} \,>\,+\nonumber \\
&+& \,2\,\Le K_{a a_4} \Ra_{- -}
<\,\B_{-}^{\,a_4} \B_{+}^{\,a_1} \B_{-}^{\,a_2} \B_{-}^{\,a_3} \,>\,,
\eeqar
where correspondingly:
\beqar\label{D2}
\D^{2}_{\bot}\,<\,\B_{-}^{\,a} \B_{-}^{\,a_1} \B_{-}^{\,a_2} \B_{+}^{\,a_3} \,>\, &= &\,-\,\imath\,\delta^{\,a a_3}\,<\,\B_{-}^{\,a_1} \B_{-}^{a_2} \,>\,+\,
\Le K_{\,a_4}^{a} \Ra^{+}_{-} <\,\B_{-}^{\,a_4} \B_{-}^{\,a_1} \B_{-}^{\,a_2} \B_{+}^{\,a_3} \,>\,+\nonumber \\
&+&\, 2\,\Le K^{a a_4} \Ra^{+ +}\,<\,\B_{+}^{\,a_4} \B_{-}^{\,a_1} \B_{-}^{\,a_2} \B_{+}^{\,a_3} \,>\,.
\eeqar
Solving these equations perturbatively and using \eq{B5}-\eq{B6} we obtain:
\beq\label{D3}
<\B_{+}^{\,a} \B_{+}^{\,a_1} \B_{-}^{\,a_2} \B_{-}^{\,a_3} > = -\,2\,\imath\,G_{0\,-}^{\,a a_{3}\,+}\,<\B_{+}^{\,a_1} \B_{-}^{a_2} >=
-2\,G_{0\,-}^{\,a a_{3}\,+}\,G_{0\,-}^{\,a_1 a_{2}\,+}- 2\,\imath\,
G_{0\,-}^{\,a a_{3}\,+}\,<\B_{+}^{\,a_1} \B_{-}^{a_2} >_1\,,
\eeq
and
\beqar\label{D4}
&\,&<\B_{-}^{\,a} \B_{-}^{\,a_1} \B_{-}^{\,a_2} \B_{+}^{\,a_3}>= -3 \imath G_{0\,-}^{\,a_1 b\,+}\,\Le K^{b b_1} \Ra^{+ +}
G_{0\,-}^{\,b_1 a_2\,+} <\B_{+}^{\,a_3} \B_{-}^{a} > - \nonumber \\ 
&-&3 \imath G_{0\,-}^{\,a_1 a_3\,+}\,\Le K^{b b_1} \Ra^{+ +}
G_{0\,-}^{\,b_1 a_2\,+} <\B_{+}^{\,b} \B_{-}^{a} >
=
-6 G_{0\,-}^{\,a a_3\,+}G_{0\,-}^{\,a_1 b\,+}\Le K^{b b_1} \Ra^{+ +}
G_{0\,-}^{\,b_1 a_2\,+}.
\eeqar

\begin{figure}[!h]
\centering
\includegraphics[scale=0.6]{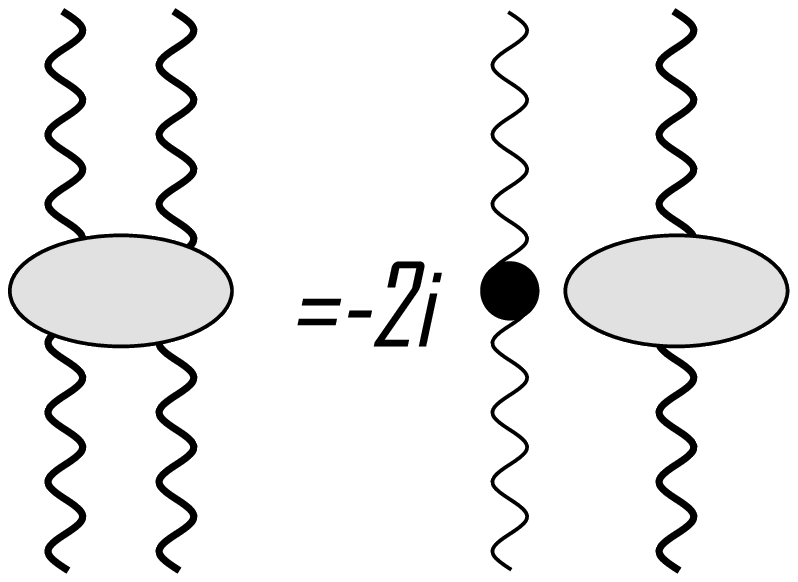}
\caption{The diagram represents \eq{D3}. It denotes LO expression of $<\B_{+}^{\,a} \B_{+}^{a_1} \B_{-}^{\,a_2} \B_{-}^{\,a_3}>$ correlator in terms of Reggeon propagator.}
\label{fig:digraph6}
\end{figure}
\begin{figure}[!h]
\centering
\includegraphics[scale=0.6]{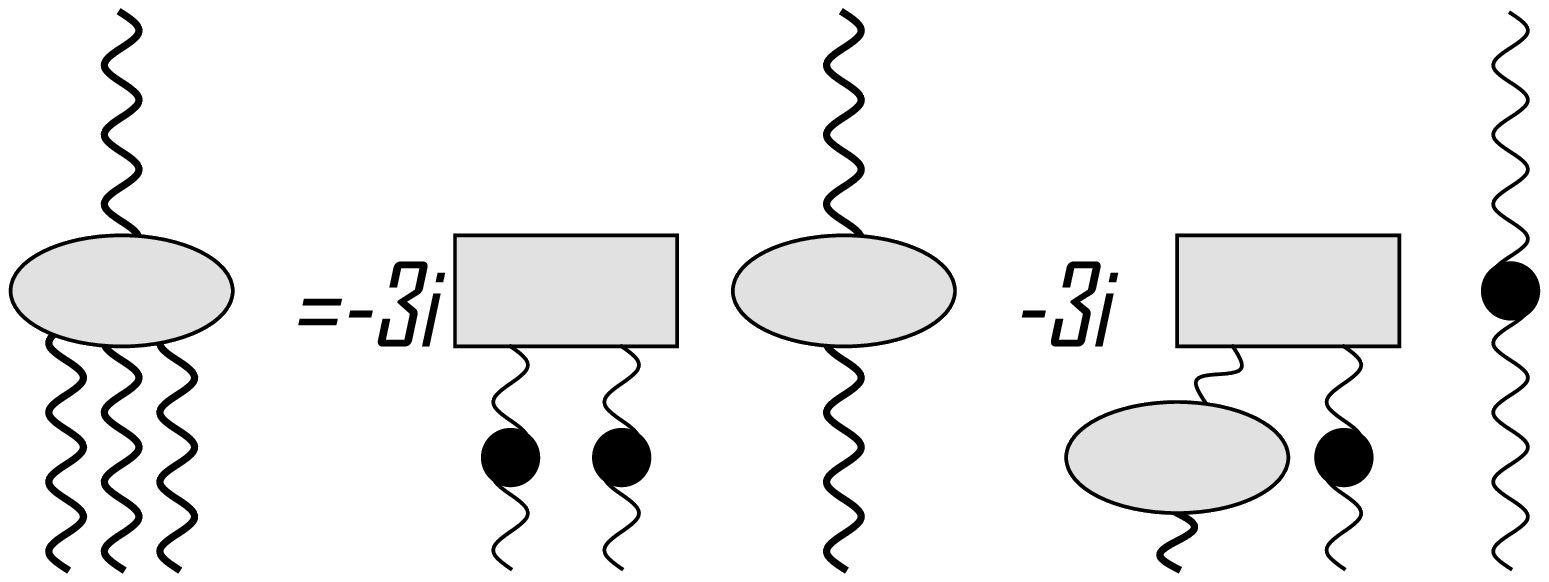}
\caption{The diagram represents  \eq{D4}. It denotes LO expression of  $<\B_{-}^{\,a} \B_{-}^{\,a_1} \B_{-}^{\,a_2} \B_{+}^{\,a_3}>$ correlator in terms of Reggeon propagator.}
\label{fig:digraph7}
\end{figure}

We note, that the leading contribution of \eq{D4} correlator is to $g^2$  order, i.e. a leading order of the $K^{+ +}$ vertex  is $g^2$ at least.
Corresponding $<\B_- \B_- \B_{+} \B_{+}>$ and $<\B_+ \B_+ \B_+ \B_->$ correlators can be obtained from the \eq{D3}-\eq{D4} expressions by the change $+$ to $-$
and vise versa in the effective vertices in the expressions.

 For the other correlators we correspondingly have:
\beqar\label{D5}
\D^{2}_{\bot}\,<\,\B_{+}^{\,a} \B_{-}^{\,a_1} \B_{-}^{\,a_2} \B_{-}^{\,a_3} \,>\, &= &\,-\,3\,\imath\,\delta^{\,a a_1}\,<\,\B_{-}^{\,a_2} \B_{-}^{a_3} \,>\,+\,
\Le K^{\,a_4}_{\,a} \Ra^{+}_{-} <\,\B_{+}^{\,a_4} \B_{-}^{\,a_1} \B_{-}^{\,a_2} \B_{-}^{\,a_3} \,>\,+\nonumber \\
&+& \,2\,\Le K_{a a_4} \Ra_{- -}
<\,\B_{-}^{\,a_4} \B_{-}^{\,a_1} \B_{-}^{\,a_2} \B_{-}^{\,a_3} \,>\,
\eeqar
and
\beq\label{D6}
\D^{2}_{\bot} < \B_{-}^{\,a} \B_{-}^{\,a_1} \B_{-}^{\,a_2} \B_{-}^{\,a_3} >=
\Le K_{\,a_4}^{\,a} \Ra^{+}_{-} <\B_{-}^{\,a_4} \B_{-}^{\,a_1} \B_{-}^{\,a_2} \B_{-}^{\,a_3} > +
2 \Le K^{\,a a_4} \Ra^{+ +} <\B_{+}^{\,a_4} \B_{-}^{\,a_1} \B_{-}^{\,a_2} \B_{-}^{\,a_3} >\,.
\eeq
Solving the system perturbatively, we obtain:
\beqar\label{D7}
< \B_{-}^{\,a} \B_{-}^{\,a_1} \B_{-}^{\,a_2} \B_{-}^{\,a_3} >\,& = &2\,G_{0\,-}^{\,a b\,+}\,
\Le K^{\,b b_1} \Ra^{+ +} <\B_{+}^{\,b_1} \B_{-}^{\,a_1} \B_{-}^{\,a_2} \B_{-}^{\,a_3} >\,=\,\nonumber \\
\,&=&\,-\,12\,\imath\,G_{0\,-}^{\,a b\,+}\,\Le K^{\,b b_1} \Ra^{+ +}\,<B_{+}^{b_1} B_{-}^{a_1}>\,G_{0\,-}^{\,a_2 c\,+}\,\Le K^{\,c c_1} \Ra^{+ +}\,G_{0\,-}^{\,c_1 a_3\,+}\,
\eeqar
that provides for the first correlator:
\beq\label{D8}
< \B_{+}^{\,a} \B_{-}^{\,a_1} \B_{-}^{\,a_2} \B_{-}^{\,a_3} > = -6 \Big[
\Le \delta - 4 G_{0\,-}^{\,+} K_{- - } G_{0\,-}^{\,+} K^{+ +} \Ra^{\,a}_{b}\Big]^{-1}\,G_{0\,-}^{\,b a_1\,+}\,
G_{0\,-}^{\,a_2 b_2\,+}\,\Le K^{\,b_2\,b_3} \Ra^{+ +}\,G_{0\,-}^{\,b_3 a_3\,+}\,,
\eeq
or to leading approximation
\beq\label{D9}
< \B_{+}^{\,a} \B_{-}^{\,a_1} \B_{-}^{\,a_2} \B_{-}^{\,a_3} >\, =\, -6 \,
G_{0\,-}^{\,a a_1\,+}\,
G_{0\,-}^{\,a_2 b\,+}\,\Le K^{\,b\,b_1} \Ra^{+ +}\,G_{0\,-}^{\,b_1 a_3\,+}\,,
\eeq
that precisely reproduce \eq{D4} expression. Hence we obtain for \eq{D7}:
\beq\label{D71}
< \B_{-}^{\,a} \B_{-}^{\,a_1} \B_{-}^{\,a_2} \B_{-}^{\,a_3} >\,=\,-12\,
G_{0\,-}^{\,a b\,+}\,\Le K^{\,b b_1} \Ra^{+ +}\,G_{0\,-}^{\,b_1 a_1\,+}\,
G_{0\,-}^{\,a_2 b_2\,+}\,\Le K^{\,b_2\,b_3} \Ra^{+ +}\,G_{0\,-}^{\,b_3 a_3\,+}\,.
\eeq

\begin{figure}[!h]
\centering
\includegraphics[scale=0.8]{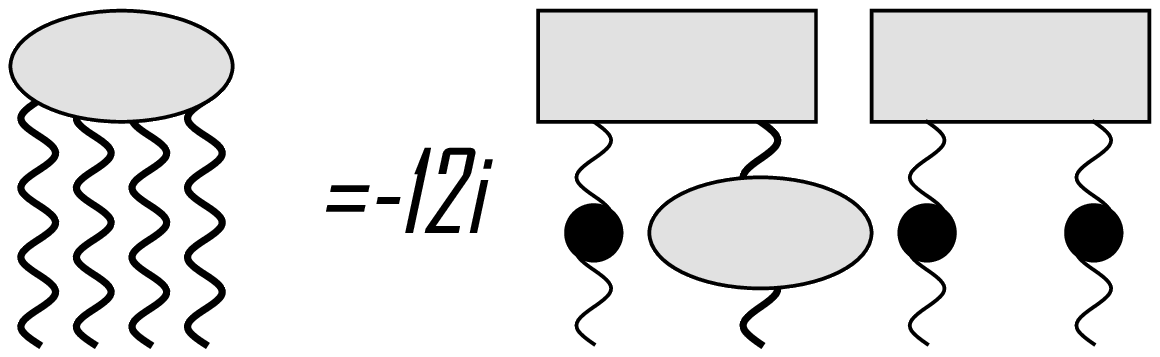}
\caption{The diagram represents \eq{D71}. It denotes $< \B_{-}^{\,a} \B_{-}^{\,a_1} \B_{-}^{\,a_2} \B_{-}^{\,a_3} >\,$ correlator in terms of Reggeon propagator.}
\label{fig:digraph8}
\end{figure}

\end{document}